\documentclass{aa} 

\usepackage{graphicx}
\usepackage[comma,authoryear]{natbib}
\usepackage{longtable}
\usepackage{amssymb}
\usepackage[varg]{txfonts}
\usepackage[usenames]{color}

\usepackage{hyperref}

\newcommand{\logg}{ {\rm log\,{\it g}}}
\newcommand{\teff}{$T_{\rm eff}$}
\newcommand{\feh}{{\rm [M/H]}}
\newcommand{\lum}{$\log(L_{\star}/{\rm L}_\odot)$}

\newcommand{\msini}{$m_{\rm p}\sin{i}$}

\newcommand{\Lstar}{L$_{\star}$}
\newcommand{\Rstar}{R$_{\star}$}
\newcommand{\Mstar}{M$_{\star}$}

\newcommand{\thetaLD}{$\theta_\mathrm{LD}$}
\newcommand{\thetaUD}{$\theta_\mathrm{UD}$}

\newcommand{\eg}{{e.g.}}
\newcommand{\ie}{{i.e.}}

\newcommand{\m}{$\mu$m}

\newcommand{\Mjup}{M$_{\rm Jup}$}
\newcommand{\Msun}{M$_{\sun}$}
\newcommand{\Rsun}{R$_{\sun}$}
\newcommand{\Lsun}{L$_{\sun}$}

\newcommand{\sophie}{S{\small OPHIE}}
\newcommand{\hipp}{H{\small IPPARCOS}}

\begin{document} %
\title{Constraints on HD113337 fundamental parameters and planetary system\thanks{Partly based on observations made with the VEGA/CHARA spectro-interferometer.}}
\subtitle{Combining long-base visible interferometry, disk imaging and high-contrast imaging}

\author{
S.~Borgniet \inst{1,2} 
\and K.~Perraut \inst{2} 
\and K.~Su \inst{3} 
\and M.~Bonnefoy \inst{2,4} 
\and P.~Delorme \inst{2} 
\and A.-M.~Lagrange \inst{2}
\and V.~Bailey \inst{5}
\and E.~Buenzli \inst{4}
\and D.~Defr\`{e}re \inst{6}
\and T.~Henning \inst{4}
\and P.~Hinz \inst{3}
\and J.~Leisenring \inst{3}
\and N.~Meunier \inst{2}
\and D.~Mourard \inst{7}
\and N.~Nardetto \inst{7}
\and A.~Skemer \inst{3}
}
\institute{
LESIA, Observatoire de Paris, Universit\'{e} PSL, CNRS, Sorbonne Universit\'{e}, Univ. Paris Diderot, Sorbonne Paris Cit\'{e}, 5 place Jules Janssen, 92195 Meudon, France
\and
Univ. Grenoble Alpes, IPAG, CNRS, F-38000 Grenoble, France
\and 
Steward Observatory, University of Arizona, 933 N Cherry Ave, Tucson, AZ 85721, USA
\and
Max Planck Institute for Astronomy, Koenigstuhl 17, D-69117 Heidelberg, Germany
\and
Jet Propulsion Laboratory, California Institute of Technology
\and
Space sciences, Technologies \&~Astrophysics Research (STAR) Institute, University of Li\`{e}ge, Li\`{e}ge, Belgium
\and
Universit\'{e} C\^{o}te d'Azur, OCA, CNRS, Lagrange, Parc Valrose, B\^{a}t. Fizeau, 06108 Nice cedex 02, France
}

\offprints{simon.borgniet@obspm.fr} \date{Received
         ...; accepted ...}

\abstract
{HD113337 is a Main-Sequence F6V field star more massive than the Sun. It hosts one confirmed giant planet and possibly a second, candidate one, detected by radial velocities. It also hosts a cold debris disk detected through the presence of an infrared excess, making it an interesting system to explore.}
{We aim at bringing new constraints on the star's fundamental parameters, the debris disk properties and the planetary companion(s) by combining complementary techniques.}
{We used the VEGA interferometer on the CHARA array to measure HD113337 angular diameter. We derived its linear radius using the parallax from Gaia Second Data Release. We computed the bolometric flux to derive its effective temperature and luminosity, and we estimated its mass and age using evolutionary tracks. Then, we used Herschel images to partially resolve the outer debris disk and estimate its extension and inclination. Next, we acquired high-contrast images of HD113337 with the LBTI to probe the $\sim$10-80 au separation range. Finally, we combined the deduced contrast maps with previous radial velocity (RV) of the star using the \texttt{MESS2} software to bring upper mass limits on possible companions at all separations up to 80 au. We took advantage of the constraints on the age and inclination (brought by the fundamental parameter analysis and the disk imaging, respectively) for this analysis.}
{We derive a limb-darkened angular diameter of $0.386 \pm 0.009$ mas that converts into a linear radius of $1.50 \pm 0.04$ \Rsun~for HD113337. The fundamental parameter analysis leads to an effective temperature of $6774 \pm 125$~K, and to two possible age solutions: one young within 14-21~Myr and one old within 0.8-1.7~Gyr. We partially resolve the known outer debris disk and model its emission. Our best solution corresponds to a radius of $85 \pm 20$ au, an extension of $30 \pm 20$ au and an inclination within 10-30\degr~for the outer disk. The combination of imaging contrast limits, published RV, and age and inclination solutions allow us to derive a first possible estimation of the true masses of the planetary companions: $\sim 7_{-2}^{+4}$~\Mjup~for HD113337~b (confirmed companion), and $\sim 16_{-3}^{+10}$~\Mjup~for HD113337~c (candidate companion). We also constrain possible additional companions at larger separations.}
{}


\authorrunning {S. Borgniet et al.}
\titlerunning{Constraints on the HD113337 system.}
\maketitle

\section{Introduction}

Thousands of exoplanets have been discovered for more than twenty years, exhibiting a wide diversity of properties (mass, separation, eccentricity, etc). 
Each planet detection method allows to estimate only some of these parameters. Most of the planetary companions known so far have been detected indirectly, either with the transit or the radial velocity (RV) method. For both techniques, the main derived parameter (\ie, the planet radius from transits, and the planet minimal mass from RV) depends directly on the values of the star parameters (\ie, the stellar radius and mass, respectively). Hence a better precision on these stellar parameters leads to better estimations of the planetary parameters \citep[][]{ligi12,stassun17,white18}. In the case of transiting companions, the combined use of the stellar mass and radius even allows to determine the true companion mass and thus, its density and expected composition \citep[][]{ligi16,crida18}.

In the case of RV planetary systems, another major uncertainty lies in the inclination ($i$) between the orbital plane and the observer line-of-sight. Indeed, RV alone only provide the companion minimal mass (\msini) and not its true mass, which is decisive to infer the companion true nature. While astrometric measurements of the primary star used in combination with RV can help to derive the true companion mass for the most massive ones \citep[brown dwarfs; see \eg~][]{sahlmann11,bouchy16}, one can also try to estimate the system inclination by looking at other proxies. Deriving the inclination of the star's rotation axis or the inclination of a resolved debris disk are two such possibilities, if assuming that they all rotate in the same plane (\ie~the orbital plane).\\

Estimating precisely the main stellar fundamental parameters, \ie~the stellar mass (\Mstar), the stellar radius (\Rstar), and the stellar age is far from being straightforward. Most estimates of these parameters are based on the use of evolutionary models with constraints brought by various observations. A direct and accurate way to obtain stellar radii is to use long-baseline interferometry to directly measure the stellar angular diameter, which allows to reach an unbiased precision of $\sim$3\%~on \Rstar~\citep[see \eg][]{ligi12,boyajian15,ligi16}. When combined to the stellar bolometric flux and the parallax, such a measurement of the stellar radius allows to derive new (and potentially unbiased) estimations of the stellar luminosity (\Lstar) and effective temperature (\teff). Then, once placed on an Hertzprung-Russell diagram, the mass and age can be determined through the use of stellar evolutionary models and the interpolation of isochrones. A good knowledge of the stellar age is essential in the case of directly imaged substellar companions, as the companion mass is determined through the use of mass-luminosity model. Hence, understanding the true nature and the formation processes of such imaged companions strongly depend on a good estimation of the age of the primary star. Famous cases are \eg~the giant planet orbiting around $\beta$~Pictoris \citep{bonnefoy14} or the companion to GJ504 \citep{dorazi17,bonnefoy18}.\\

A key challenge to develop the theory of planetary formation and evolution processes is to understand the respective influence of the different stellar characteristics (\eg~the stellar mass, metallicity, effective temperature, etc) on these processes. While stellar metallicity is well known to be positively correlated to the giant planet (GP) frequency \citep{fischer05}, a correlation between the stellar mass and the GP frequency and/or mass is yet to be fully investigated for stars more massive than the Sun \citep[see \eg~][and references therein]{borgniet17}. Interactions between giant planets and debris disks are another key topic to investigate in the context of planetary evolution.\\

We make here a case study of a system of high interest. \object{HD113337} is a Main-Sequence star more massive than the Sun, that hosts one (possibly two) RV-detected giant planet(s) \citep[][hereafter BO19+]{borgniet18}, as well as an unresolved debris disk. We use multi-technique observations to better understand and/or constrain the properties of the primary star (through optical interferometry), to resolve the debris disk, and to explore the system's outer environment (through deep imaging). This paper is structured as follows: in Sect.~\ref{sect:system}, we present the HD113337 system, looking at the star, the debris disk and the planetary system. Second, we review in Sect.~\ref{sect:obs} the different observations that we made and the data reduction processes that we used. We present and discuss our results in Sect.~\ref{sect:results}. We specifically show how the combination of these different techniques allow us to better understand and constrain the HD113337 system.

\section{The HD113337 system}\label{sect:system}

\subsection{The star}

HD113337 is a Main-Sequence F6V-type star located at a distance $d = 36.2 \pm 0.2$~pc from the Sun \citep[based on the parallax given by the {\it Gaia} second Data Release or DR2,][and see details in Sect.~\ref{sect:param}]{gaia16,GDR2}.
Its stellar mass is consistently estimated to be $\sim$1.4~\Msun~in the literature, based on spectroscopic \citep[][hereafter AP99+]{allende99} or photometric analyses \citep[Geneva-Copenhagen Survey -- hereafter GCS III+,][]{casagrande11}. Its estimated effective temperature ranges from $6670 \pm 80$ K based on the photometry (GCS III+) to $6760 \pm 160$~K based on the spectroscopy (AP99+). The fit of the spectral energy distribution (SED) by \cite{rhee07} gives \teff~$= 7200$~K and provides a stellar radius estimation of $1.5 \pm 0.15$~\Rsun. As it is a field dwarf star, the age of HD113337 is the most difficult stellar parameter to estimate. The typical isochronal age derived from the photometric \teff~is $1.5_{-0.55}^{+0.43}$~Gyr (GCS III+). It is in agreement with the age of $1.6_{-0.8}^{+2.2}$~Gyr derived by \cite{david15} from Str\"{o}mgren photometry. We conducted two different analyses to derive independently an age estimation of HD113337 in \cite{borgniet14}. Briefly, we first estimated the age of the bound distant (projected separation of $\sim$120~as or $\sim$4400~au) M-type companion \object{2M1301+6337} to be $100_{-50}^{+100}$~Myr. Second, we measured HD113337 Lithium abundance and estimated the corresponding age to be $> 160$~Myr, leaving our analysis unconclusive \citep[for more details, see][and references therein]{borgniet14}. Finally, activity- and rotation-related age diagnostics such as the relations derived by \cite{mamajek08} do not apply to such an early spectral type. 

\subsection{The debris disk}

HD113337 exhibits a clear infrared (IR) excess from $\sim$20~\m~up to 1200~\m~with a $L_{\rm IR}/L_{\rm BOL} = 10^{-4}$ fractional luminosity. Based on data from the {\it InfraRed Astronomical Satellite} (IRAS), \cite{rhee07} estimated the dust temperature to be $\sim$100~K and radius to be 18~au. Using {\it Spitzer} data that provide a better coverage on the spectral energy distribution (SED) of the disk, \cite{moor11} concluded that the disk has a dust temperature of $\sim$53~K, suggesting a disk radius of $55 \pm 3$~au. A more recent study by \cite{chen14} found that HD113337 SED was best fitted by a two-belt model, with a first, warm ($316 \pm 10$~K) dust ring located at 1.7~au from the star and a second, cold ($54 \pm 5$~K) dust ring at 179~au. The system was observed with the James Clerk Maxwell Telescope (JCMT) for the SCUBA-2 Observations of Nearby Stars (SONS) survey at both 450~\m~and 850~\m~by \cite{holland17}, who reported upper limits of $<$75 mJy (5$\sigma$) and $<$3.6 mJy (3$\sigma$), respectively.

\renewcommand{\arraystretch}{1.25}
\begin{table*}[t!]
\caption{HD113337 VEGA observation and reduction log. Only the selected $V^{2}$ measurements (see text) are displayed.}
\label{tab:V2}
\begin{center}
\begin{tabular}{l l c c c c c c c c c c}\\
\hline
\hline
  Date     & Calibrator&$\theta_{\rm UD}^{\rm CAL}$&$\sigma_{\theta}^{\rm CAL}$  & UT        &Calibration &$\lambda_{\rm obs}$ & Baseline & $B_{\rm P}$ & P.A. & $V^{2}$& $r_{0}$ \\
           & (HD)      &  (mas)                & (mas)                   &(on target)& sequence   &   (nm)           &          &  (m)       & (\degr)&     &(cm)    \\
\hline
2013-05-25 & 111270 (C1)& $0.307$ & $0.021$      & 04:35     & C1-S-C1    & 730              & E2W1     &  250   & -131   & $0.321 \pm 0.040$& 9 \\
           & 110462 (C2)& $0.221$ & $0.014$      &           &            & 710              & E2W1     &  250   & -131.5 & $0.357 \pm 0.046$& \\
           &            &         &              & 06:23     & C1-S-C2    & 750              & E2W1     &  249   & -100   & $0.420 \pm 0.048$& \\
           &            &         &              &           &            & 730              & E2W1     &  249   & -100   & $0.323 \pm 0.032$& \\
           &            &         &              &           &            & 710              & E1E2     &  64    & -121   & $0.818 \pm 0.052$& \\
           &            &         &              &           &            & 710              & E2W1     &  249   & -100   & $0.362 \pm 0.042$& \\
\hline
2014-07-03 & 98772      & $0.230$ & $0.014$      & 04:32     & C-S-C      & 710              & E2W2     &  156   & -176   & $0.780 \pm 0.058$& 8 \\
           &            &         &              &           &            & 700              & E2W2     &  156   & -176   & $0.708 \pm 0.053$& \\
\hline
2014-07-07 & 98772      & $0.230$ & $0.014$      & 04:17     & C-S-C      & 710              & E2W2     &  156   & 174    & $0.650 \pm 0.051$& 6 \\
           &            &         &              &           &            & 700              & E2W2     &  156   & 174    & $0.648 \pm 0.052$& \\
\hline      
2015-05-30 & 118214     & $0.230$ & $0.015$      & 06:01     & C-S-C      & 710              & E2W2     &  156   & -141   & $0.712 \pm 0.023$& 7 \\
           &            &         &              &           &            & 700              & E2W2     &  156   & -141   & $0.658 \pm 0.020$& \\
\hline
2015-06-01 & 121409     & $0.226$ & $0.015$      & 06:02     & C-S-C      & 710              & E1E2     &  66    & -151   & $0.914 \pm 0.018$& 10\\
           &            &         &              &           &            & 710              & E2W1     & 245    & -130   & $0.510 \pm 0.070$& \\
           &            &         &              &           &            & 700              & E1E2     &  66    & -151   & $0.902 \pm 0.017$& \\
           &            &         &              & 06:45     & C-S-C      & 710              & E1E2     &  66    & -148   & $0.940 \pm 0.018$& \\
           &            &         &              &           &            & 700              & E1E2     &  66    & -148   & $0.918 \pm 0.017$& \\
\hline 
\hline
\end{tabular}
\end{center}
Columns 1 and 5 give the observation date and UT time (on the science target). Columns 2, 3 and 4 give the calibrator identifier in the HD catalog, its uniform-disk angular diameter ($\theta_{\rm UD}^{\rm CAL}$) in the $R$-band, and the 1$\sigma$ error bar on the calibrator diameter ($\sigma_{\theta}^{\rm CAL}$). We took the calibrator diameters from the JMMC Stellar Diameters Catalog Version 2 \citep[JSDC,][]{bourges17}, while we kept the corresponding 1$\sigma$~uncertainties from the JSDC Version 1 \citep[][see text]{lafrasse10}. Column 6 gives the calibration sequence followed for the observation and column 8 is the baseline used. Column 7 gives the central wavelength of the 20 nm-wide spectral bands in which we computed the $V^{2}$. Columns 9 and 10 give the projected base length $B_{\rm P}$ and its orientation PA. Column 11 gives the corresponding calibrated $V^{2}$ value. Column 12 gives the Fried parameter (estimation of the quality of the atmosphere) for each observation night. Note that three observation points are not displayed here as the results were fully rejected (see text).
\end{table*}
\renewcommand{\arraystretch}{1}

\subsection{The planetary system}

From 2006 to 2016, a RV survey of 125 northern AF-type dwarf stars (including HD113337) was carried out with the \sophie~spectrograph at Observatoire de Haute-Provence (France). The aim was to search for giant planets and brown dwarfs (BD) around Main-Sequence stars more massive than the Sun. Clear periodic variations of HD113337 RV were detected and attributed to the presence of a $\sim$3~\Mjup~GP orbiting around the star with a $\sim$320-day period \citep[$\sim$1 au,][]{borgniet14}. After monitoring the system with \sophie~for three additional years, HD113337 RV were found to exhibit a second periodicity on a longer timescale. The possible sources for this RV long-term variability were investigated and it was concluded that a possible origin was the presence of a second GP with a \msini~$\sim$7~\Mjup~minimal mass on a wider orbit ($\sim$5~au, see BO19+).\\

The combined presence of a debris disk, giant planet(s), as well as an ill-constrained age, makes HD113337 an object of high interest.

\section{Observations and data processing}\label{sect:obs}

\subsection{Optical interferometry of HD113337}

The {\it Center for High Angular Resolution Astronomy} array \citep[hereafter CHARA,][]{chara} is the main optical and near-infrared interferometric array in the northern hemisphere. It hosts six 1-m telescopes arranged by pairs in a Y shape and oriented to the west (W1 and W2), east (E1 and E2) and south (S1 and S2), allowing a wide range of baseline orientations. The corresponding baselines range from $\sim$30~m to $\sim$330~m (\ie~a maximal angular resolution of 0.2-0.3~mas in the visible). The Visible SpEctroGraph and polArimeter \citep[hereafter VEGA,][]{mourard09} is one of the instruments operating in the visible at the focus of the CHARA array. VEGA is a spectro-interferometer which allows to combine the light coming from 2 to 4 telescopes simultaneously, at different spectral resolutions ($R$ = 6000 and 30000).\\ 

We observed HD113337 with two different telescope triplets (E1E2W2 and E1E2W1) chosen to (partially) resolve its small expected angular diameter \citep[$\sim$0.4~mas given its $\sim$27~mas parallax and the 1.5~\Rsun~radius from][]{rhee07}. For each observation point, we tried to follow a calibrator-target-calibrator sequence (C-S-C, see Table~\ref{tab:V2}) with either 30 or 40 blocks of 2500 short (10~ms) exposures per star to ensure an instrumental transfer function stable enough to correctly calibrate the target squared visibilities ($V^{2}$). While it is not necessarily mandatory, observing the calibrator (C) star twice (\eg~before and after the science (S) target) allows to monitor and take into account possible variations of the transfer function during the observation time \citep[see][and below]{mourard12}. Furthermore and if possible, using two different calibrators with well-defined angular diameters (C$_{\rm 1}$-S-C$_{\rm 2}$ sequence) instead of one (C-S-C sequence) reinforces the robustness of the target $V^{2}$ computation by reducing its dependency to the calibrator diameter value and uncertainty. We used the SearchCal software \citep{bonneau06} to select adequate calibrators at the different observation epochs. We acquired ten observation points on five different nights from May 2013 to June 2015, using VEGA red spectral channel in the $\sim$700 to $\sim$750~nm range. We computed the $V^{2}$ values in either two or three 20~nm-wide spectral bands (depending on the observing conditions) with the standard VEGA reduction pipeline $vegadrs$ \citep{mourard12}. Due to the small angular diameters of both our target and calibrators, we had to discard a significant part of the data that revealed themselves not robust enough. We first discarded three entire observation points (UT~06:37 on 2013-05-25, UT~04:59 on 2014-07-07 and UT~07:32 on 2015-06-01), for which either the calibrated target $V^{2}$ computation process did not converge or the signal-to-noise ratio (S/N) on the $V^{2}$ values was very low ($\simeq$0). It also happens that we were not able to obtain measurements on the second calibrator for these three observation points (C-S sequence), which may have hindered the estimation of the transfer function at the time of the observations. Second, we also discarded $V^{2}$ measurements with S/N~$<$4 \citep{mourard12}, while ensuring that this did not bias our results \citep[as done by][]{perraut13}. The detail of our observations (after selection) is given in Table~\ref{tab:V2}.

\begin{figure}[t!]
\centering
\includegraphics[width=1.\hsize]{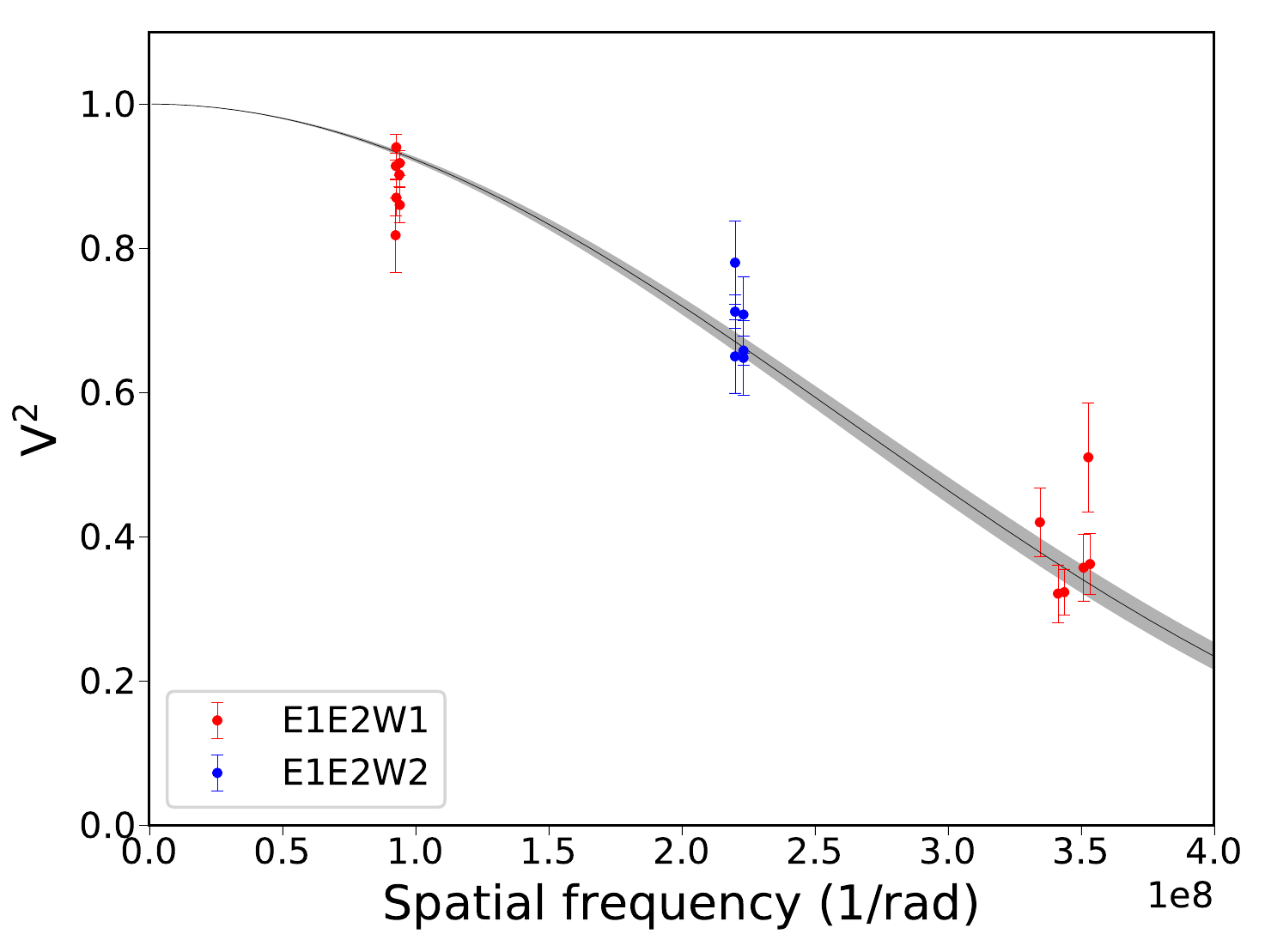}
\caption{VEGA squared visibilities versus spatial frequency for HD113337. The $V^{2}$ are displayed as red or blue circles (depending on the telescope triplet). The solid line and gray zone represent the uniform-disk LITpro best model along with its uncertainty.}
\label{fig:V2}
\end{figure}

The JMMC Stellar Diameter Catalog (JSDC) that provides the calibrator angular diameters has recently been updated \citep{bourges17}, with a significant increase (by 6-12\%~here) on the diameters of early-type calibrators ($\theta_{\rm UD}^{\rm CAL}$) and diameter uncertainties ($\sigma_{\theta}^{\rm CAL}$) smaller by $\sim$50\%. Here we chose to use the $\theta_{\rm UD}^{\rm CAL}$ values from the more recent JSDC2 while conservatively keeping the $\sigma_{\theta}^{\rm CAL}$ values from the JSDC1 (Table~\ref{tab:V2}). The $\theta_{\rm UD}^{\rm CAL}$ change from JSDC1 to JSDC2 translates into an average increase of $\sim$7\%~on our $V^{2}$ values lower than 0.6 (2013-05-25 data) and no significant change on our $V^{2}$ values above 0.6. We fitted the visibility measurements with the JMMC fitting engine LITpro \citep{LITPRO} based on a modified Levenberg-Marquardt algorithm\footnote{\url{www.jmmc.fr/litpro\_page.htm}}, and derived a uniform-disk (UD) angular diameter \thetaUD~$= 0.371 \pm 0.009$~mas. We display the best model of the visibility function derived with LITpro in Fig.~\ref{fig:V2}. If using the $\theta_{\rm UD}^{\rm CAL}$ values from the JSDC1 instead, we obtain \thetaUD~$= 0.364 \pm 0.009$~mas (\ie~less than a 2\%~difference). Furthermore, using the smaller $\sigma_{\theta}^{\rm CAL}$ values from JSDC2 would lead to a \thetaUD~uncertainty of 0.007~mas instead of 0.009~mas. Our \thetaUD~robustness is due to the strong constraints brought on our model by the five measurements made with the E2W1 intermediate baseline on 2013-05-25 (our best data, see Fig.~\ref{fig:V2}). 

We used the linear limb-darkening coefficients in the $R$-band provided by \citet{Claret2011} to derive the corresponding limb-darkened angular diameter \thetaLD. For a solar metallicity and a null microturbulent velocity, we computed the limb-darkened diameters for \teff~in the 6500~to 7000~K range, and \logg~in the 4 to 4.5 range. These coefficients vary at the third decimal level within the considered \teff~and \logg~ranges. We thus considered the average limb-darkening coefficient on our parameter space and obtained a limb-darkened angular diameter of \thetaLD~$= 0.386 \pm 0.009$~mas (hence a $\sim$2.4\%~precision).\\

\begin{figure}
\centering
\includegraphics[width=1.\hsize]{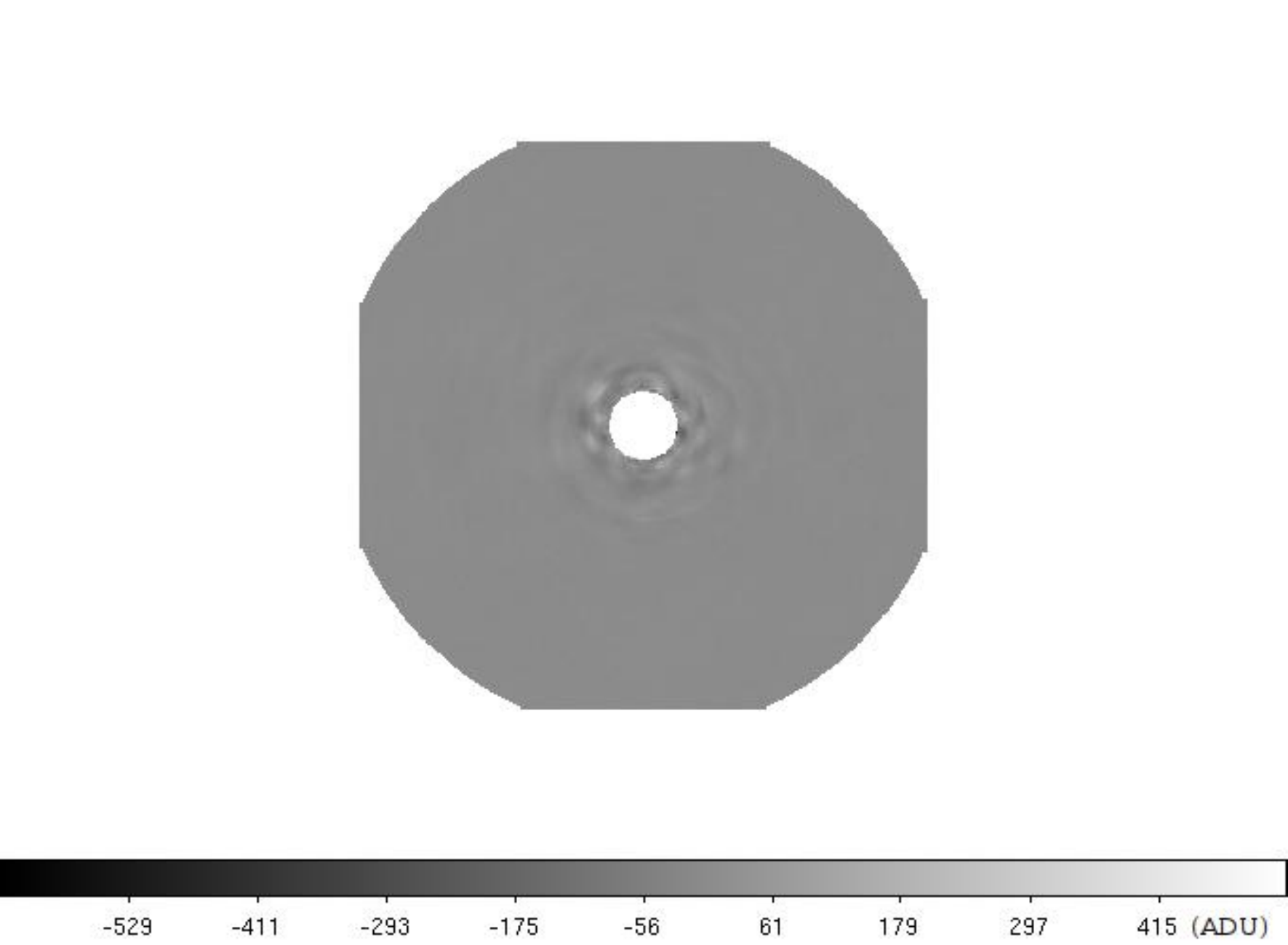}
\caption{LBTI (left eye) image of HD113337. North is up and East is left. The size of the image is $\sim$3.2 arcsecond.}
\label{fig:lbti_image}
\end{figure}

\subsection{High-contrast imaging}\label{sect:lbti}

HD113337 was observed on January 7, 2015 with the LMIRCam near-infrared camera \citep{hinz08,skrutskie10} at the {\it Large Binocular Telescope Interferometer} (LBTI). The LBTI was operated in double-aperture mode: the secondary deformable mirrors were used to record two side-by-side adaptive-optics (AO) images of HD113337 recorded by LMIRCam at $L$-band (3.68 - 3.88 \m). The telescope+instrument do not have a derotator. Therefore, it automatically enables for passive angular differential imaging \citep[ADI,][]{marois06}. We obtained 597$\times$4.95 second and 605$\times$4.95 second AO exposures of the target for the right and left eyes of the telescope, respectively. Less AO exposures were obtained on the right eye because of open loops. The field orientation changed by $35.1$\degr~and $37.2$\degr~during that sequence of exposures on the right and left eyes, respectively. The core of the star's point-spread function was saturated over a diameter of $\sim$128 mas during the observations to increase the dynamics of the recorded images. We therefore had to acquire non-satured exposures of the star before and after the sequence of saturated exposures for calibrating the astrometry and photometry using a neutral density (attenuation factor of $9\times10^{-3}$).  

\begin{figure*}
\centering
\includegraphics[width=1.\hsize]{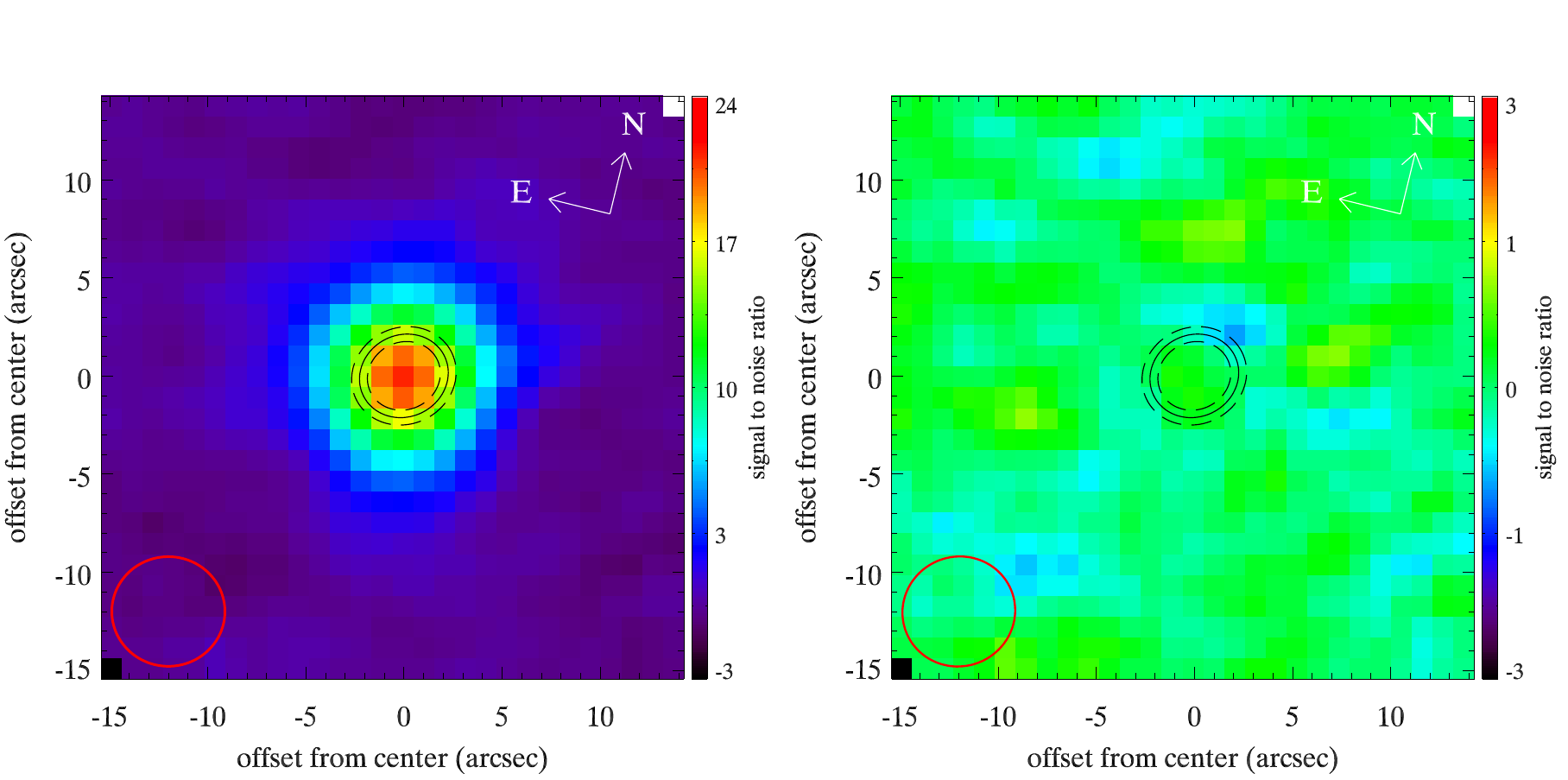}
\caption{Left: Herschel 70~\m~image of the HD113337 system. The best-fit disk can be described as a Gaussian ring peaked at 85~au (solid ellipse) with a width of 30~au (marked by dashed ellipses), viewed at 25\degr~from face-on. The PACS 70~\m~beam is shown as the red ellipse on the lower left corner. Right: the residual image after the subtraction of the best-fit model.}
\label{fig:pacsim}
\end{figure*} 

The data were reduced using the MPIA ADI pipeline \citep{bonnefoy14b}. The pipeline carried out the basic cosmetic steps on the raw frames (de-trending of the raw frames, bad pixel interpolation, sky-background subtraction, flat field calibration). The star position was registered in the resulting frames using the \texttt{mpfit2Dpeak.pro} IDL function\footnote{\url{http://www.physics.wisc.edu/~craigm/idl/fitting.html}} which allowed for a bi-dimensional Moffat function to be fitted onto the PSF wings while masking the saturated core. The parallactic angles were computed at the time of the observations. We applied the LOCI algorithm \citep{lafreniere07} to evaluate and subtract the stellar halo in the left and right eye data, independently. No point source could be detected following that step (Fig.~\ref{fig:lbti_image}).

\subsection{Herschel observations of the cold outer disk}

The Herschel data of HD113337 were obtained using the {\it Photodetector Array Camera and Spectrograph} (PACS) instrument \citep{poglitsch10}, in the mini scanmap mode with simultaneous observations at 70 and 160 \m~under the open time program $OT2\_ksu\_3$ (Fig.~\ref{fig:pacsim}). The data were obtained on March 3, 2012 with OBSID~1342243344 and 1342243345. Herschel PACS data reduction were performed following the procedure published by \cite{balog14} for calibration stars.

We fitted a 2-D Gaussian function to estimate the source's Full Width at Half Maximum (FWHM). At 70~\m, the measured FWHM is 7\farcs44$\times$7\farcs15, $\sim$1.3 times larger than the FWHM of typical point sources (5\farcs76$\times$5\farcs58). The measured FWHM at 160~\m~is slightly larger than the typical value for point sources, but less significant. Because the source is marginally resolved at PACS wavelengths, we used aperture sizes of 12" and 22" to measure photometry with a sky annulus of 35"-- 45" at 70~\m~and 160~\m, respectively. Including 7\%~of absolute flux calibration, the final PACS fluxes are: 177.1$\pm$12.5~mJy and 118.5$\pm$9.3~mJy at 70~\m~and 160~\m, respectively. The PACS 70~\m~flux agrees very well with the previously published {\it Multiband Imaging Photometer for Spitzer} (MIPS) 70~\m~photometry \citep[][see below]{moor11}.

\section{Results}\label{sect:results}

\subsection{Determination of the stellar fundamental parameters}\label{sect:param}

\paragraph{Linear radius --}

We used the {\it Gaia} \citep{gaia16} DR2 parallax $\pi_{\rm P} = 27.61 \pm 0.04$~mas \citep{GDR2}. According to the {\it Gaia} documentation, the published DR2 parallax uncertainties may be underestimated by a factor of $\sim$10\%~for bright stars such as HD113337 ($G \simeq 5.9$~mag). Furthermore, there are potential systematic errors on the DR2 parallaxes such as a global zero-point offset \citep{lindegren18}. We used the formula given by \cite{lindegren18} for bright stars to recompute the error on HD113337 DR2 parallax: \ie~we quadratically summed the published parallax uncertainty scaled by a factor 1.08 with an additional uncertainty of 0.021 mas. We thus obtained $\pi_{\rm P} = 27.61 \pm 0.05$~mas. The corresponding distance is $36.2 \pm 0.2$~pc, \ie~in good agreement with the HD113337 distance of $36.18 \pm 0.06$~pc derived by \cite{bailer-jones18} based on a Bayesian analysis of {\it Gaia} DR2 parallaxes.\\

We used our limb-darkened angular diameter \thetaLD~in combination with the above {\it Gaia} parallax to derive the stellar radius and its error through a Monte-Carlo simulation
\begin{equation}
    R_{\star} \pm \delta R_{\star} = \frac{\theta_{\rm LD} + \delta \theta_{\rm LD}}{9.305 \times (\pi_{\rm P} + \delta \pi_{\rm P})}.
\end{equation}
We obtained \Rstar~=~$1.50 \pm 0.04$~\Rsun~(precision better than 2.7\%).

\begin{figure*}[ht!]
\includegraphics[width=1\hsize]{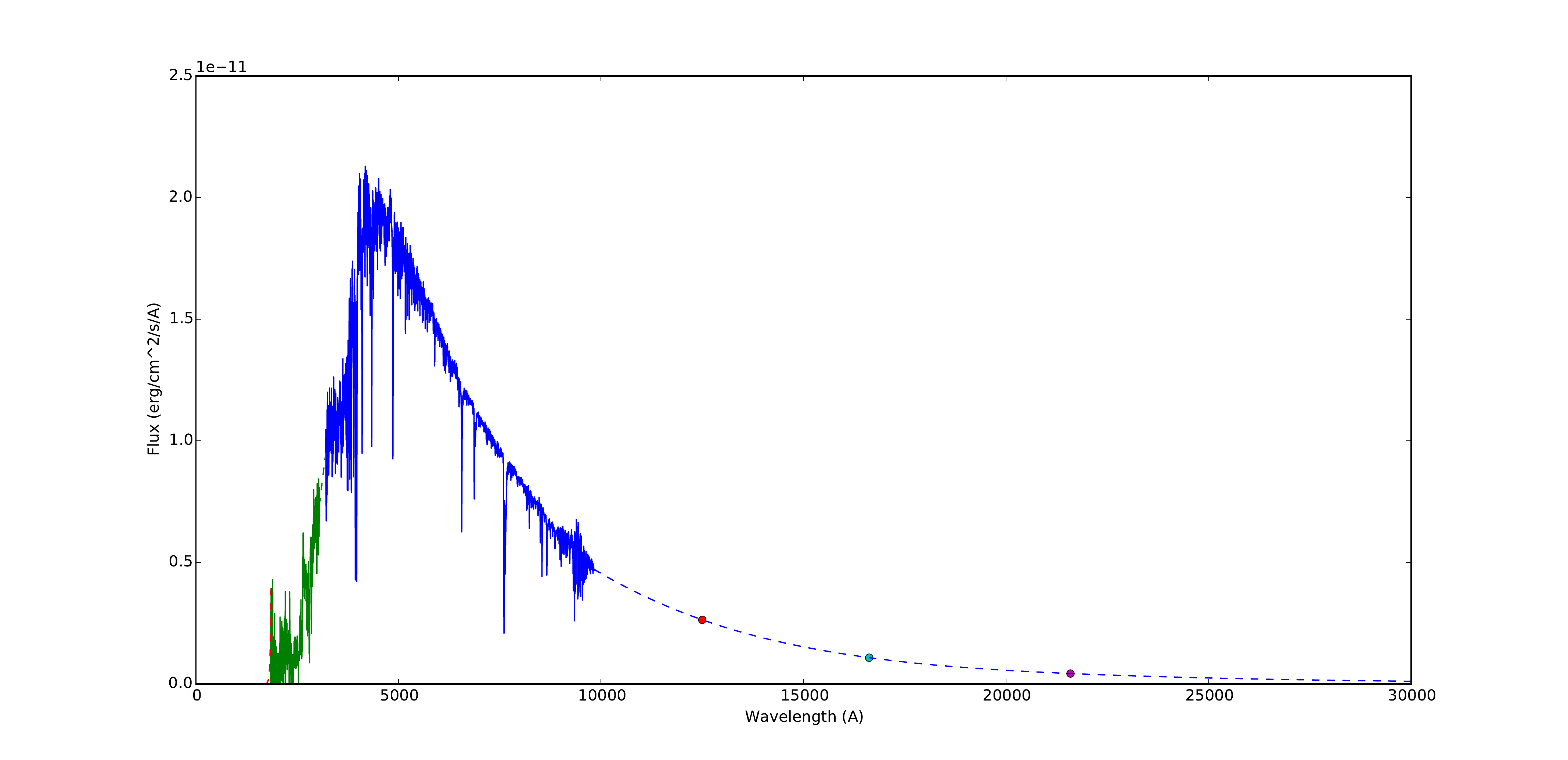}
\caption{Bolometric flux computation: INES spectrum (green), STELIB spectrum (blue), 2MASS magnitudes (circles), and interpolations (dash lines).}
\label{fig:bolom_flux}
\end{figure*}

\paragraph{Bolometric flux --}

We collected spectroscopic and photometric data to compute the bolometric flux $f_{bol}$. We determined HD113337 SED, and then computed the area under the curve of that distribution (Fig.~\ref{fig:bolom_flux}). The flux distribution of HD113337 was determined by concatenating several flux distributions:
\begin{itemize}
\item for the ultraviolet region, the rebinned spectrum from the Sky Survey Telescope obtained by the IUE \lq Newly Extracted Spectra\rq (INES) data archive\footnote{\url{http://sdc.cab.inta-csic.es/cgi-ines/IUEdbsMY}}. Based on the quality flag listed in the IUE spectra \citep{gar97}, we removed all bad pixels from the data as well as measurements with negative flux;
\item for the visible and red regions, the STELIB spectrum \citep{leborgne03};
\item in the near-infrared range, the J, H, and K 2MASS magnitudes. We used the formula in \cite{cohen03} to convert 2MASS magnitudes in fluxes.
\end{itemize}
At the shortest wavelengths, we performed a linear interpolation on logarithmic scale between 912~\AA~and 1842~\AA, considering zero flux at 912~\AA. At the longest wavelengths, we performed a linear interpolation on logarithmic scale using the 2MASS magnitudes and assuming zero flux at $1.6 \times 10^6$~\AA. We estimated the uncertainty associated to the bolometric flux by considering the following conservative uncertainties, \ie, 3\%~uncertainty on the flux computed from the STELIB spectrum, 10\%~on the flux computed from the combined IUE spectra, and 15\%~on the flux derived from interpolations. Finally, we obtained a bolometric flux $f_{\rm bol} = 1.05 \pm 0.06 \times 10^{-7}$~erg/cm$^2$/s.

\paragraph{Luminosity and effective temperature --}

From the parallax and the bolometric flux we derived the luminosity through a Monte Carlo method
\begin{equation}
L_{\star} = 4 \pi \, f_{\rm bol} \, {C^2} / {\pi_{\rm p}^2},
\end{equation}
where $C$ is the conversion from parsecs to cm ($3.086 \times 10^{18}$), and $\pi_{\rm p}$ the parallax in arcseconds. We found \Lstar~=~$4.29 \pm 0.25$~\Lsun, where the error bar is dominated by that of the bolometric flux.

We finally derived the effective temperature \teff~from \thetaLD~and $f_{\rm bol}$ through a Monte Carlo method 
\begin{equation}
\sigma T_{\rm eff}^{4} = f_{\rm bol} \, \Big( C \times 9.305 / (\theta_{\mathrm{LD}} \times R_{\sun}) \Big)^{2},
\end{equation}
where $\sigma$ stands for the Stefan-Boltzmann constant ($5.67 \times 10^{-5}$~erg/cm$^2$/s/K$^{-4}$). We determined \teff~$= 6774 \pm 125$~K.

\paragraph{Position in the Hertzsprung-Russell diagram --}

\begin{figure*}[ht!]
\centering
\includegraphics[width=0.49\hsize]{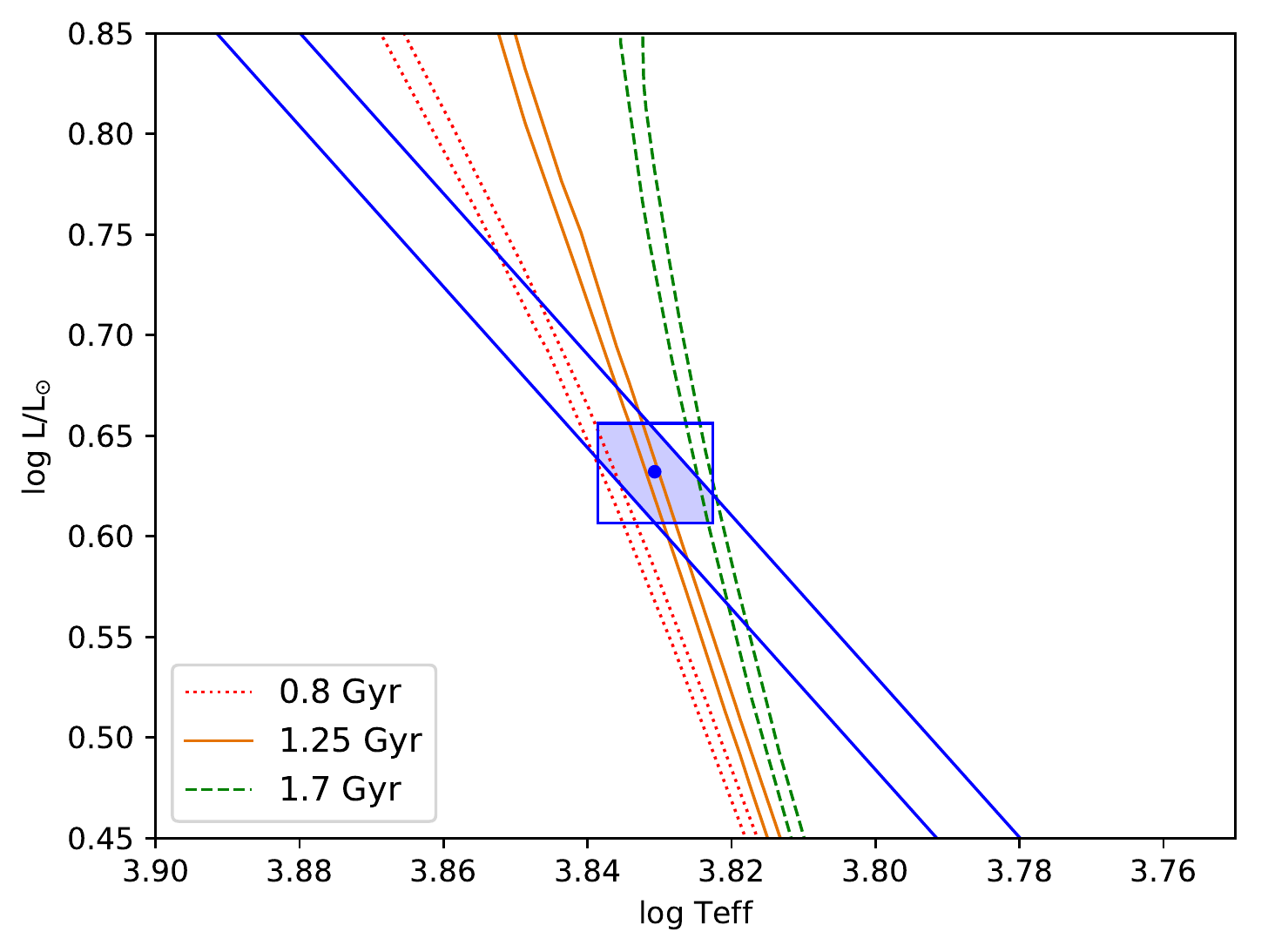}
\includegraphics[width=0.49\hsize]{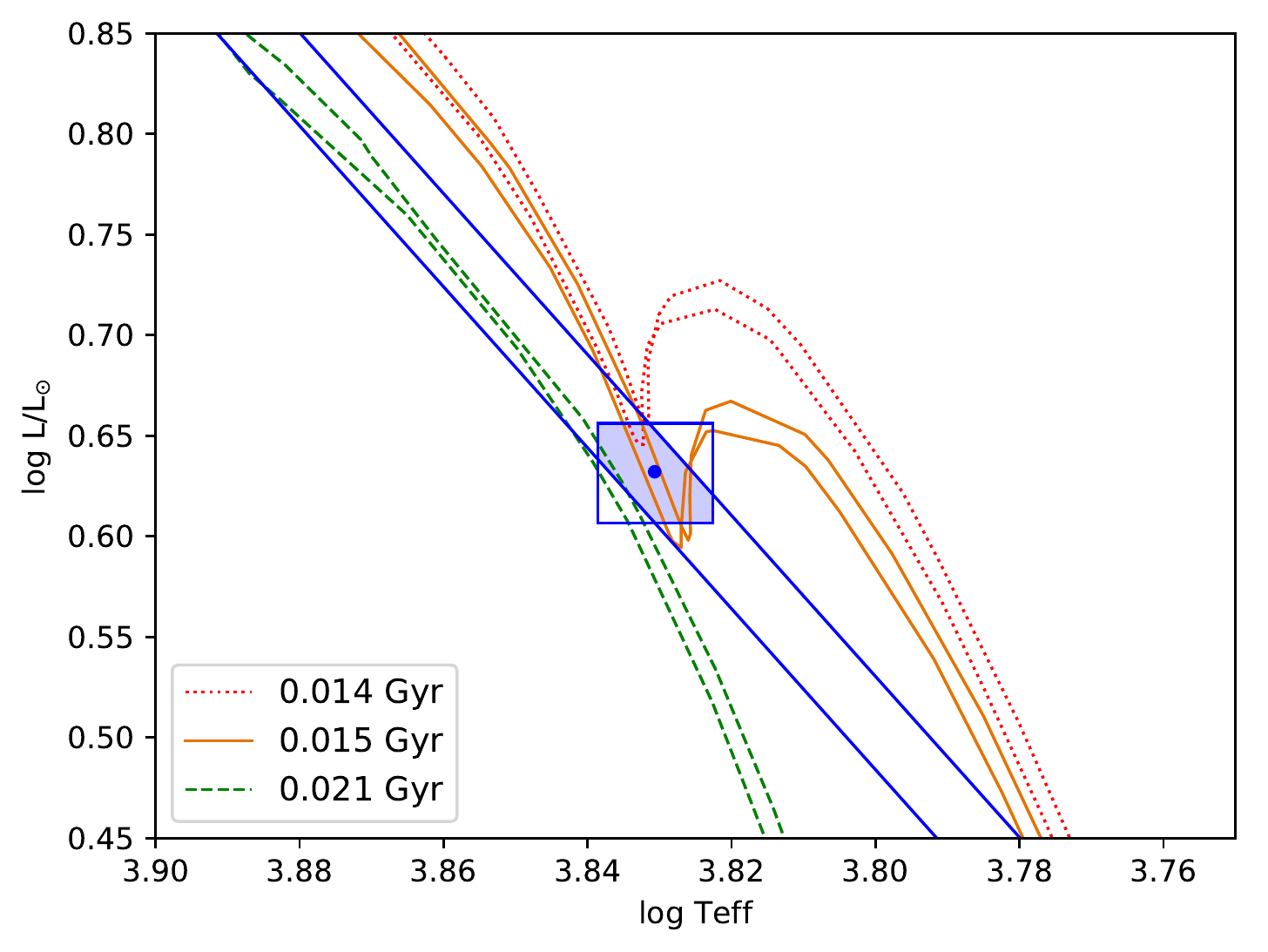}
\caption{Position of HD113337 in the Hertzprung-Russell diagram. Left plot: ``old'' solution; right plot: ``young'' solution. On both plots, the blue dot and blue box represent the [$\log(T_{\rm eff})$, \lum] values and 1$\sigma$ uncertainties (respectively) deduced from our analysis. The blue solid straight lines represent in the same way the constraints (1$\sigma$ uncertainties) on \Rstar. On both plots, evolutionary tracks for $Z$ in the range 0.0169 to 0.0185 are displayed for three cases: best age solution (solid orange curves), lower age limit (dotted red curves), and upper age limit (dashed green curves).}
\label{fig:evol_tracks}
\end{figure*}

\renewcommand{\arraystretch}{1.25}
\begin{table*}[t]
\centering
\caption{Fundamental parameters of HD113337.}
\label{tab:param}
\begin{tabular}{llccc|lr}
\hline
\hline
Parameter     & Unit                    & GCS III+~\tablefootmark{(a)}    & AP99+~\tablefootmark{(b)}     & \citet{rhee07}         & \multicolumn{2}{c}{This work} \\
\hline
\thetaLD      & [mas]                   & --                             & --                             & --                     & \multicolumn{2}{c}{$0.386 \pm 0.009$} \\
$d$           & [pc]                    & $36.9 \pm 0.4$~\tablefootmark{(c)}&$37.4 \pm 0.7$~\tablefootmark{(d)}& $37.4 \pm 0.7$~\tablefootmark{(d)} & \multicolumn{2}{c}{$36.2 \pm 0.2$~\tablefootmark{(e)}}\\
\logg      &        [cm\,s$^{\rm -2}$]               & 4.2                            & $4.21 \pm 0.08$                 & --                    & \multicolumn{2}{c}{--} \\
\feh       &        [$\rm log\,\sun$]                 &  0.09                         &    --                            &   --               & \multicolumn{2}{c}{$0.07 \pm 0.02$~\tablefootmark{(f)}}\\
\Rstar     & [\Rsun]                 & --                            &$1.55 \pm 0.07$                  &$1.50 \pm 0.15$        & \multicolumn{2}{c}{$1.50 \pm 0.04$}\\
$f_{\rm bol}$   & [10$^{-7}$ erg/cm$^2$/s] & 1                             & --                               & --                    & \multicolumn{2}{c}{$1.05 \pm 0.06$}\\
\Lstar    & [\Lsun]                 & --                            & --                               & --                   & \multicolumn{2}{c}{$4.29 \pm 0.25$}\\
\teff         & [K]                     & $6670 \pm 80$                 & $6760 \pm 160$                   & 7200                 & \multicolumn{2}{c}{$6774 \pm 125$}\\
\Mstar     & [\Msun]                 & $ 1.39 \pm 0.05$              & $1.41 \pm 0.09$                  &  --                  & $1.48 \pm 0.08$~\tablefootmark{(g)} & $1.40_{-0.05}^{+0.03}$~\tablefootmark{(h)}    \\
Age           & [Gyr]                   & $1.50_{-0.55}^{+0.43}$           & --                               &0.05~\tablefootmark{(i)} & $15_{-1}^{+6} \times 10^{-3}$~\tablefootmark{(g)} & $1.25 \pm 0.45$~\tablefootmark{(h)}\\
\hline
\hline
\end{tabular}
\tablefoot{
\tablefoottext{a}{\cite{casagrande11}}
\tablefoottext{b}{\cite{allende99}}
\tablefoottext{c}{From the parallax of \cite{vanleeuwen07}.}
\tablefoottext{d}{From the parallax of \cite{esa97}.}
\tablefoottext{e}{From the {\it Gaia} DR2 parallax, \citet{GDR2}.}
\tablefoottext{f}{From \cite{soubiran16}.}
\tablefoottext{g}{Young solution.}
\tablefoottext{h}{Old solution.}
\tablefoottext{i}{From \cite{zuckerman04}.}
}
\end{table*}
\renewcommand{\arraystretch}{1}

From the determined fundamental parameters, we set our target in the Hertzsprung-Russell diagram. We used the isochrone tool CMD~2.7\footnote{\url{http://stev.oapd.inaf.it/cgi-bin/cmd\_2.7}} to derive the mass and the age of HD113337. We considered a metallicity \feh~=~$0.07 \pm 0.02$ based on different spectroscopic analyses \citep{boesgaard86,soubiran16}, thus $Z$ spanning from 0.0169 to 0.0185, and $Y$ spanning from 0.279 to 0.282. We obtained two solutions in agreement with our 1-$\sigma$ error box: a young solution corresponding to an age of $15^{+6}_{-1}$~Myr and a mass of $1.48 \pm 0.08$~\Msun, and an old solution corresponding to an age of $1.25 \pm 0.45$~Gyr and a mass of $1.40^{+0.03}_{-0.05}$~\Msun~(Fig.~\ref{fig:evol_tracks}).\\ 

The fundamental parameters (\Rstar, \teff) that we derived are generally in close agreement with previous determinations (Table~\ref{tab:param}). Regarding the stellar age and mass, finding such a degeneracy between a young and an old solution appears to be a typical result when carrying out this approach \citep[see \eg][]{ligi16,bonnefoy18}. The mass and age from our old solution agree with the mass and age ranges determined by \citet{casagrande11} and AP99+. However, our two age solutions for the adopted metallicity range do not fit the age of 150$^{+100}_{-50}$~Myr that we adopted in \citet{borgniet14}. Waiving the age degeneracy of HD113337 remains problematic so far as we know. Even combining our interferometric radius with a technique such as asteroseismology \citep{creevey07} to derive the stellar mass would prove fruitless, as the two mass values corresponding to our two age solutions are already consistent together. There are hints that the rate of debris disks detected through the presence of an IR excess decreases with the stellar age \citep{montesinos16}, yet this does not allow us to rule out the old age solution in the specific case of HD113337.

\subsection{Outer disk geometry}

We adopted a simple approach to estimate the disk extent at 70~\m~(Fig.~\ref{fig:pacsim}, left panel) by assuming that the disk emission can be described by an axisymmetric model, like a Gaussian ring defined by the peak ($R_p$) and the width (FWHM) of the ring ($R_w$). The disk has a total flux, $F_{tot}$, at 70~\m, and its midplane is assumed to be inclined by an angle of $i$ from face-on (\ie, $i = 0$\degr), with the major axis along a position angle (P.A.). We generated a series of high-resolution model images and convolved them with the observed point spread function (PSF) derived from the calibration stars. We then determined the best-fit parameters (five free parameters) by comparing the convolved model images with the observation using a $\chi^2$ statistic. The best-fit parameters are: $R_p = 85 \pm 20$~au, $R_w = 30 \pm 20$~au, $i = 25$\degr$_{-15^{\circ}}^{+5^{\circ}}$, P.A. $=128$\degr$\pm 5$\degr, and $F_{tot} = 175 \pm 12$~mJy. The right panel of Figure~\ref{fig:pacsim} shows the image residuals, all within $\pm$1$\sigma$ after the subtraction of the best-fit model.\\

\begin{figure}[t!]
\centering
\includegraphics[width=1.\hsize]{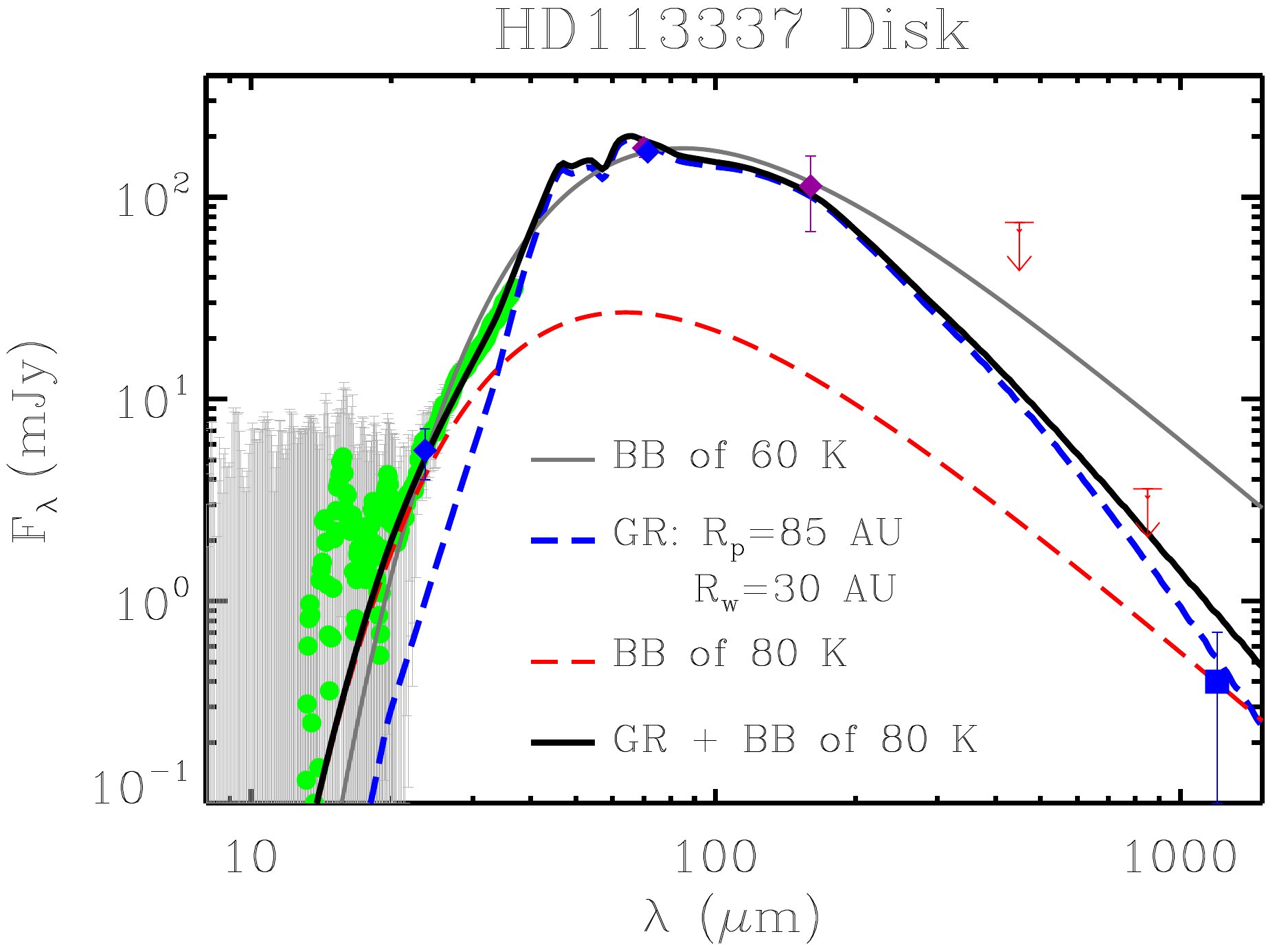}
\caption{Spectral energy distribution (SED) of the debris around HD113337, composed of broad-band photometry and mid-infrared spectrum after the removal of the stellar photosphere. Blue diamonds are the Spitzer/MIPS photometry from \cite{moor11}, green dots (with uncertainties shown in grey area) are the Spitzer/IRS spectrum from \cite{chen14}, purple diamonds are the Herschel/PACS fluxes from this study. Also plotted are the JCMT/SCUBA2 upper limits \citep{holland17} and IRAM/MAMBO2 1.2~mm from \cite{moor11}. The mid- and far-IR broad-band photometry can be described by a simple blackbody emission of 60 K (thin grey line); however, it is slightly too high compared to the IRS spectrum. The Gaussian ring (GR) SED is shown as the blue dashed line. The disk SED from $\sim$20~\m~to 1.2~mm is best described by the combination of a cold GR plus a 80~K blackbody emission (see text for details).}
\label{fig:disksed}
\end{figure} 

To make sure that the best-fit disk parameters are consistent with the observed SED, we computed the model SED using the derived geometric parameters (\ie, $R_p$ and $R_w$). Since the disk geometric parameters are derived using the PACS 70~\m~data, we only try to reproduce the SED longward of 70~\m. Assuming the disk is optically and geometrically thin and has a surface density distribution best described by the derived Gaussian ring, we are able to reproduce the bulk part of the SED using icy silicate grains \citep[the icy grain model from][for the HD95086 system]{su15}. The grain size distribution is assumed to be a power law form, $\sim a^{-3.5}$, where $a$ is the grain radius with a minimum $a_{min}$ and maximum $a_{max}$ cutoffs. We found that $a_{min}$ of $\sim$2~$\mu$m and $a_{max}$ of 1~mm can fit the far-IR SED well (Fig.~\ref{fig:disksed}). The minimum grain size (2~\m) is roughly the radiation blowout size assuming a bulk density of 1.7~g~cm$^{-3}$, a typical minimum grain size in a collisional cascade debris disks. A total dust mass for this cold disk is 7.3$\times10^{-3} M_{\rm Earth}$ (up to 1~mm grains). We note that the model cold-disk SED does not fit the MIPS 24~\m~and IRS data, which might be related to the warm component reported by \cite{chen14}. To explore this possibility, we tried to fit the mid-IR part of the SED with a simple blackbody function, and found that a blackbody emission with a temperature of 80~K represents the mid-IR SED well (Fig.~\ref{fig:disksed}). The 80~K emission is too cold compared to the warm component derived by \cite{chen14}. It might be an intermediate separate component in the disk structure, which will not be the first time that a complex disk structure is inferred \citep[\eg~$\epsilon$ Eri,][]{su17}. Alternatively, this mid-IR emission can also arise from a small amount of dragged-in grains from the cold disk under the influence of Poynting-Robertson (P-R) drag \citep[\eg][]{vanlieshout14}. Since our Herschel data do not have enough resolution to spatially resolve the inner edge of cold disk, both scenarios are possible.

\subsection{Constraints on actual and possible companions}\label{sect:mess2}

\paragraph{Known planetary system --}

We recomputed the minimal masses of the planetary companion HD113337~b and candidate c, using the new {\it Gaia} parallax, the new \Mstar~values from our fundamental parameter analysis, and our orbital analysis from BO19+. For HD113337~b we obtain \msini~$= 3.1 \pm 0.2$~\Mjup, and for the candidate companion HD113337~c we obtain \msini~$= 7.2 \pm 0.5$~\Mjup~(without significant differences between the two stellar mass solutions). Understandably, the \msini~do not change significantly with respect to the values we derived in BO19+ (\ie~$3 \pm 0.3$~\Mjup~and $6.9 \pm 0.6$~\Mjup~for HD113337~b and c, respectively), due to the little difference on the adopted stellar mass and parallax values.

\begin{figure}
\centering
\includegraphics[width=1.\hsize]{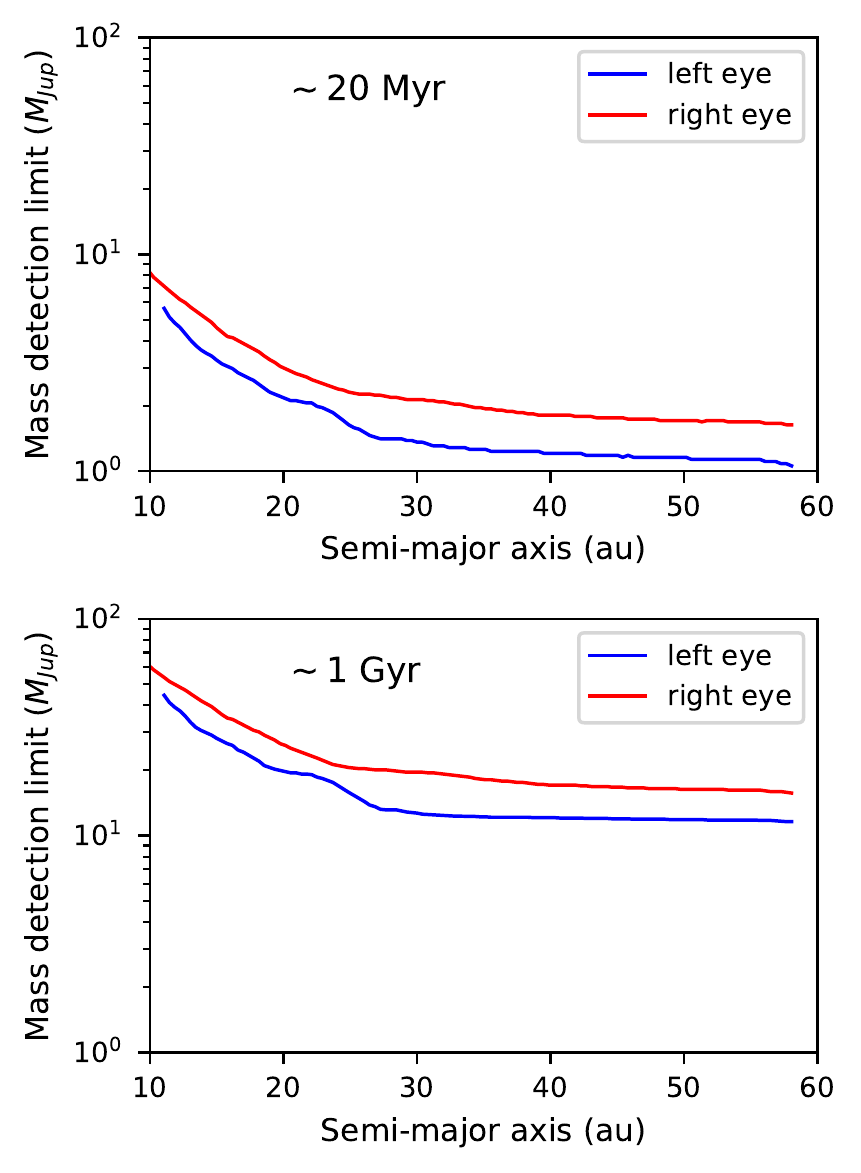}
\caption{``Classical'' mass detection limits derived from the LBTI images of HD113337, for the left (blue) and right (red) eyes. The top panel corresponds to an age of 20 Myr and the bottom panel to an age of 1 Gyr (see text).}
\label{fig:lbti_limdets}
\end{figure}

HD113337~b and candidate c orbit too close to the primary to be detected or even mass-constrained with an imager such as the LBTI. However, we can use the debris disk inclination value that we derived from the disk modeling (\ie, $i = 25$\degr$_{-15^{\circ}}^{+5^{\circ}}$) as a possible starting point. If we assume that the GP(s) and the partially resolved outer disk orbit within the same plane, we can then estimate their true mass(es). Assuming such a system inclination, the true mass of HD113337~b would then be $7.2_{-1.5}^{+4.2}$~\Mjup. The true mass of the tentative HD113337~c would be $16.4_{-3.4}^{+9.6}$~\Mjup. These true masses would be more than twice the value of the corresponding minimal masses from BO19+, in agreement with the small inclination considered here. We emphasize that this hypothesis remains widely speculative at this stage. Yet, if confirmed, this would make HD113337 GP(s) very massive planetary companions. This would be in agreement with the trend of higher GP masses with increasing stellar masses predicted by the core-accretion theory \citep{kennedy08,kennedy09}.

\begin{figure*}[ht!]
\centering
\includegraphics[width=0.49\hsize]{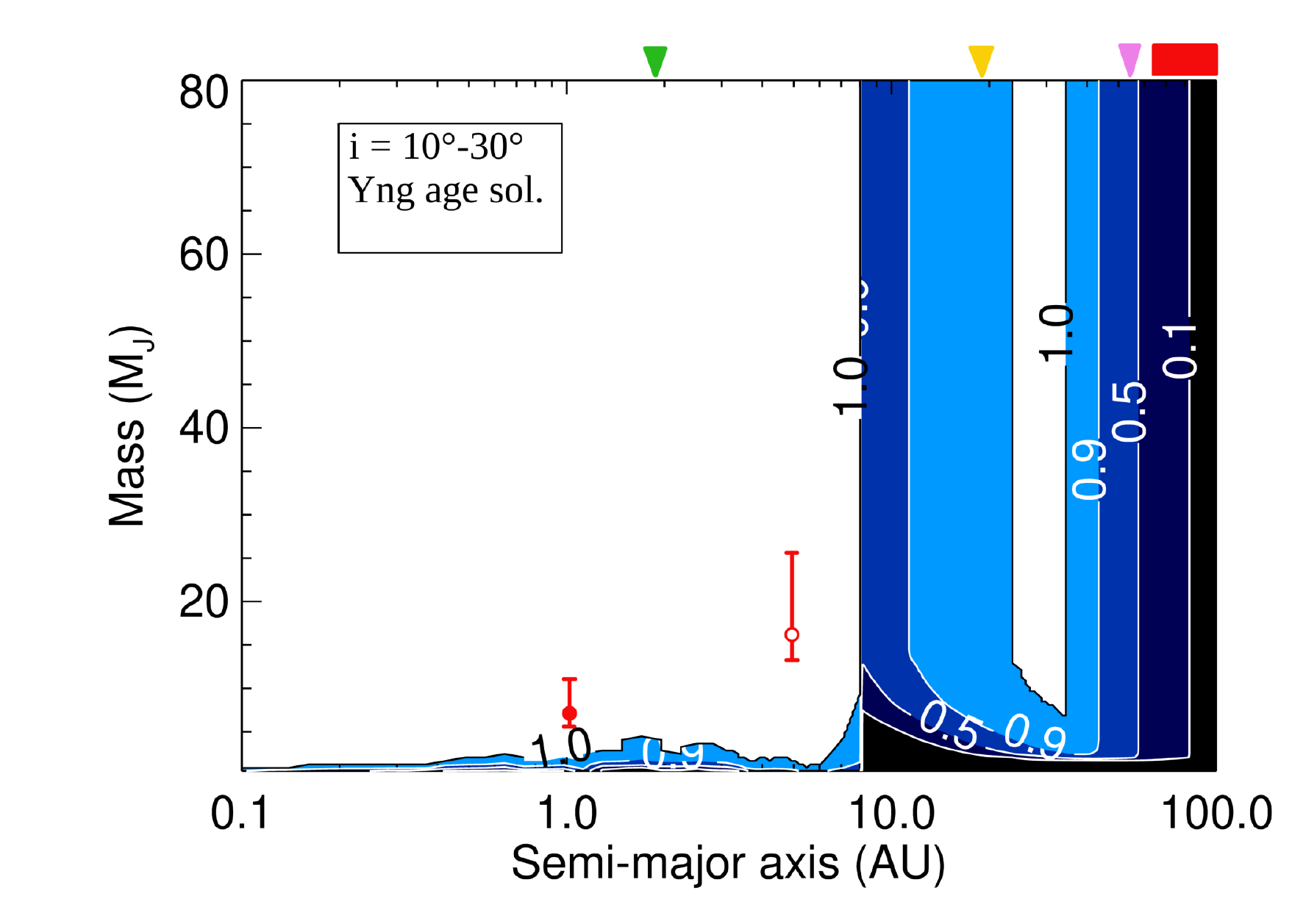}
\includegraphics[width=0.49\hsize]{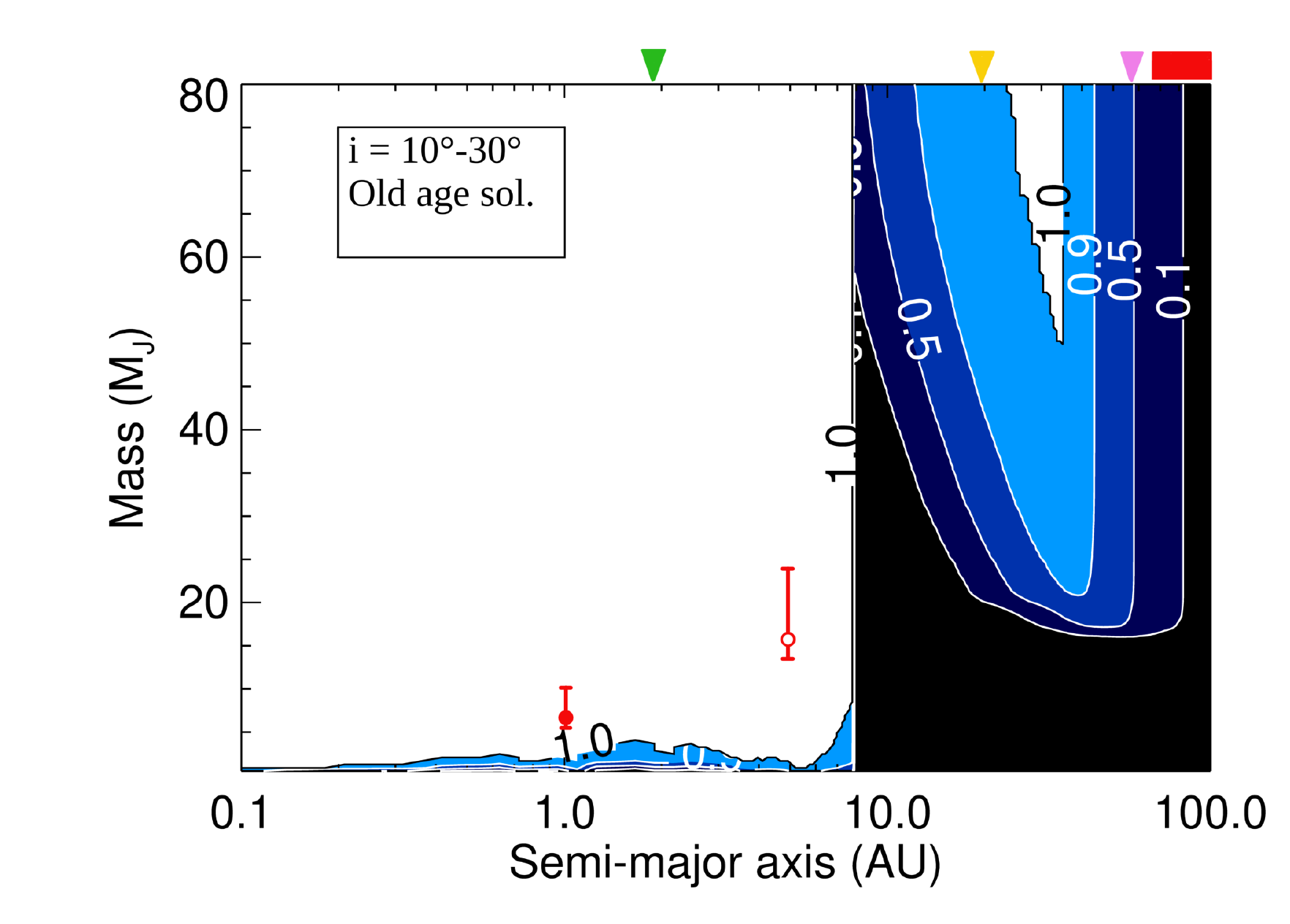}
\includegraphics[width=0.49\hsize]{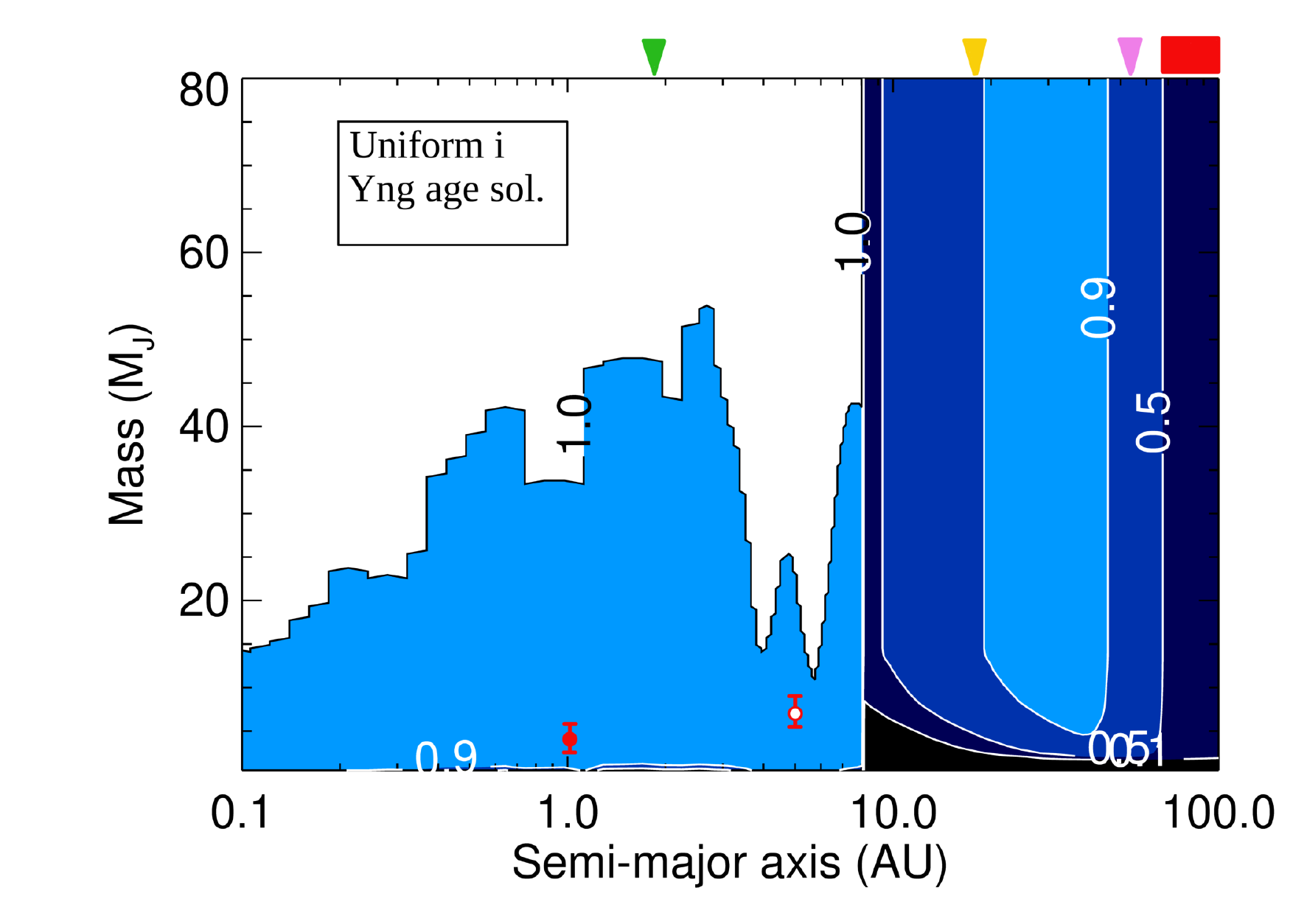}
\includegraphics[width=0.49\hsize]{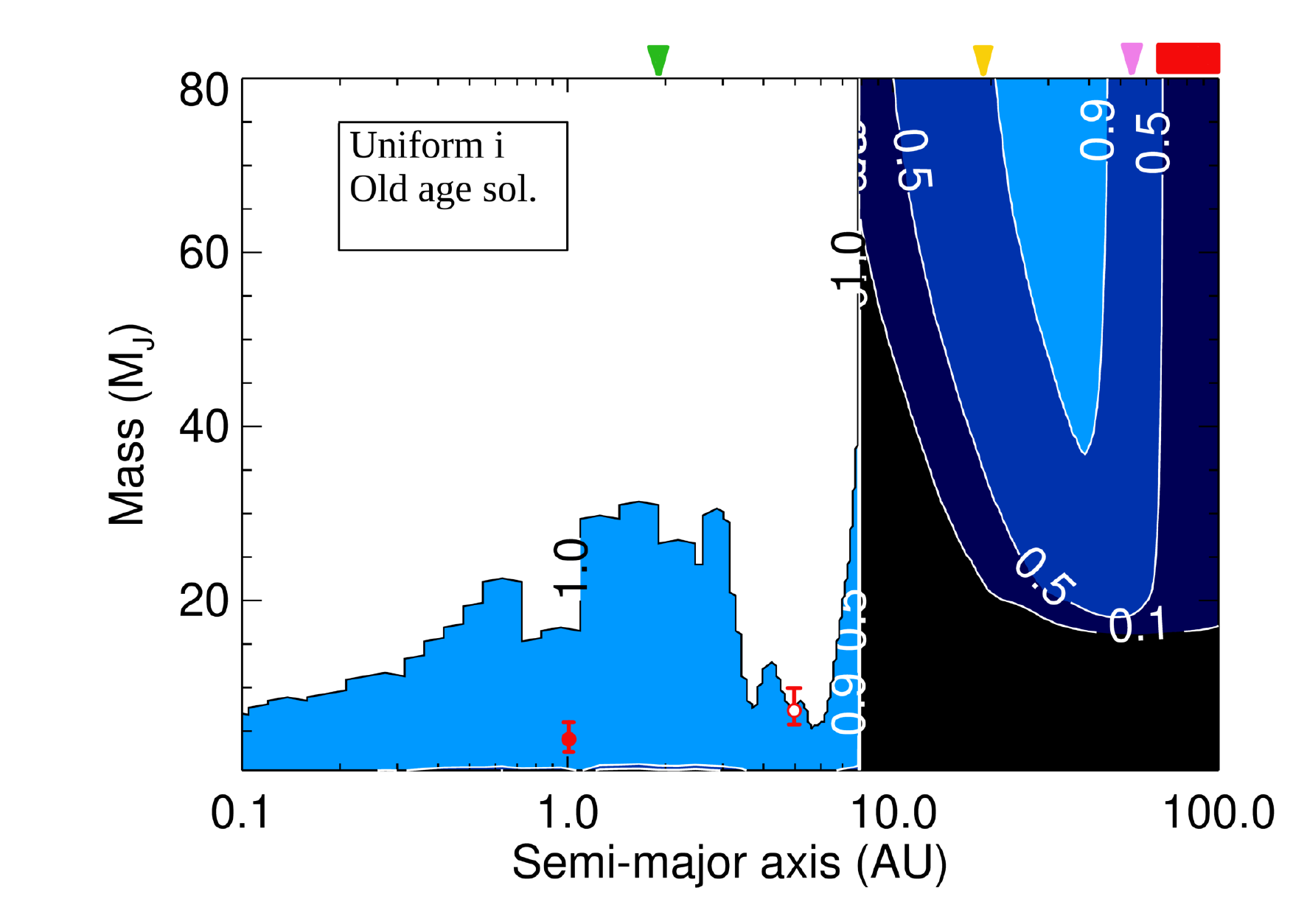}
\caption{\texttt{MESS2} detection probabilities of HD113337. The four plots correspond to our four (age; $i$) cases ($\sim$20~Myr or $\sim$1~Gyr, and $i$ within 10 to 30\degr or uniform $i$ distribution, respectively). On each plot, the contour colors (from white to black) correspond to a higher or lower (respectively) detection probability of additional companions, as indicated by the numbers: 1 indicates a detection probability over 99\%, 0.9 over 90\%, etc. We display the assumed true masses of HD113337~b and c (full and empty red dots, respectively) on the top plots (inclination assumed to be 25\degr), and their minimal masses on the bottom plots. On the top of each plot, we indicate the outer debris disk extension resolved in this study with a red band. We also indicate the disk position previously assumed from SED fits from the literature: from \cite{moor11} (purple triangle), \cite{rhee07} (yellow triangle), and the inner disk component from \cite{chen14} (green triangle), respectively.}
\label{fig:MESS2}
\end{figure*}

\paragraph{Combined mass detection probabilities --}

From the LBTI images, we estimated the flux-losses at each separation associated to the ADI process \citep{bonnefoy14} and derived detection limits (1-D) and detection maps (2-D) in contrast. We classically converted our detection limits in contrast into masses using the star distance, the WISE W1 magnitude \citep{cutri14} as a proxy of the L'-band star magnitude, and the COND tracks \citep{baraffe03}. We considered two respective ages of 20~Myr and 1~Gyr, roughly corresponding to the young and old age solutions derived from our stellar fundamental parameter analysis (Sect.~\ref{sect:param}). We display the derived mass detection limits in Fig.~\ref{fig:lbti_limdets}. The difference in terms of achieved companion sensitivity between our two age solutions (\ie~roughly one order of magnitude) highlights the importance of an accurate age determination.\\

We brought additional constraints on the mass of possible additional companions by combining our contrast detection (2-D) maps with radial velocities (RV). For this purpose, we used the {\it Multi-epoch multi-purpose Exoplanet Simulation System} \citep[\texttt{MESS2},][]{lannier17} tool. \texttt{MESS2} generates populations of synthetic planets with masses and orbital parameters within pre-defined ranges through a Monte Carlo simulation. For each of the synthetic planets, the synthetic RV signal generated at the RV observation epochs, and the simulated planet projected separation at the image's observation epoch are simultaneously compared to the RV and imaging data, respectively. With respect to ``classical'' mass detection limits derived from contrast (as in Fig.~\ref{fig:lbti_limdets}), the advantage of this approach is twofold: (1) it allows to explore different hypotheses on the companion orbital properties; and (2) it allows to assess the companion mass detection probabilities in the combined separation range covered by RV and imaging (\ie~from the star's close proximity out to the imager's field of view). \texttt{MESS2} was successfully applied to famous systems such as AU~Mic \citep{lannier17}, HD95086 \citep{chauvin18}, $\beta$~Pictoris \citep{lagrange18}, and GJ504 \citep{bonnefoy18}.

\begin{figure*}[ht!]
\centering
\includegraphics[width=0.49\hsize]{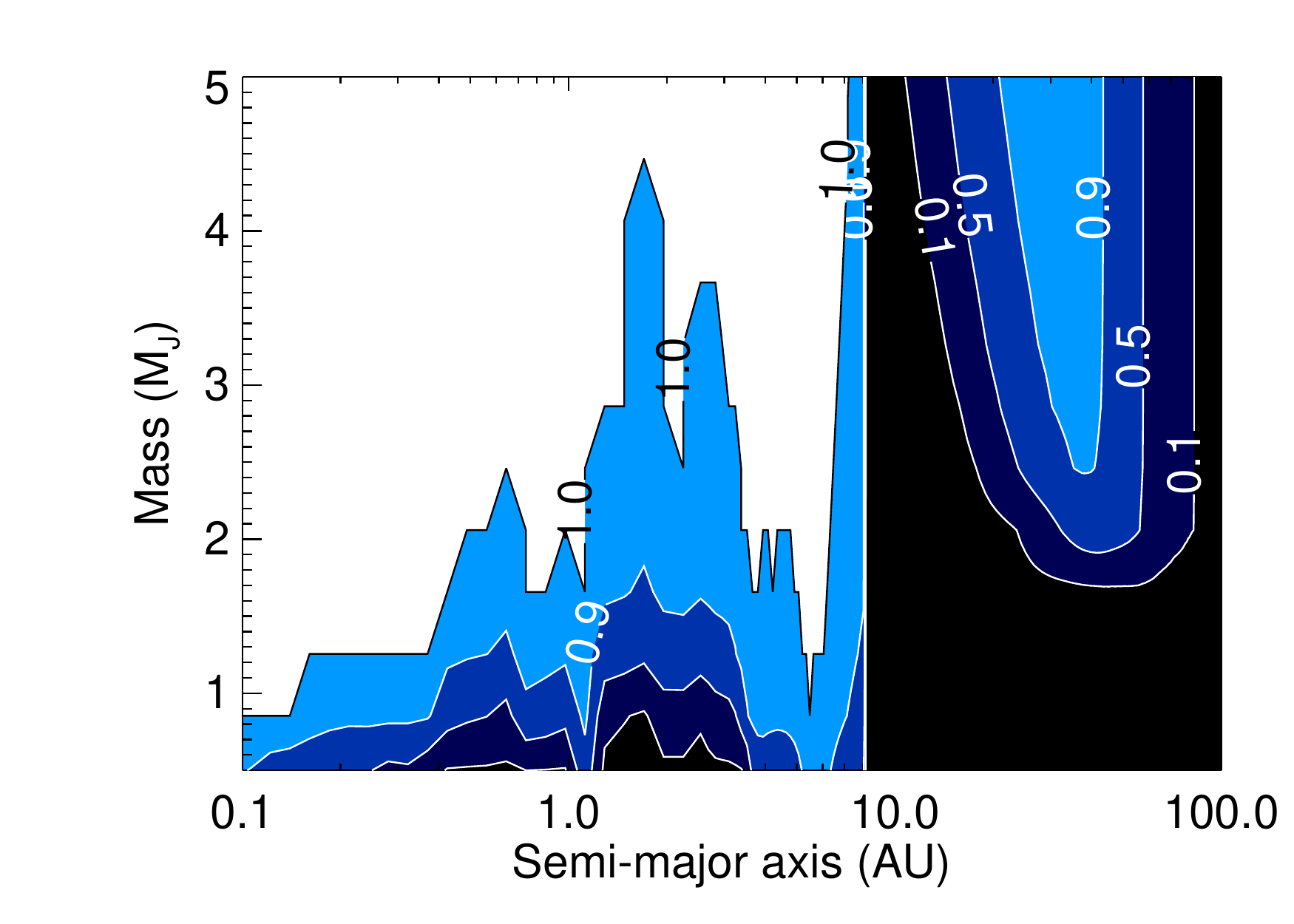}
\includegraphics[width=0.49\hsize]{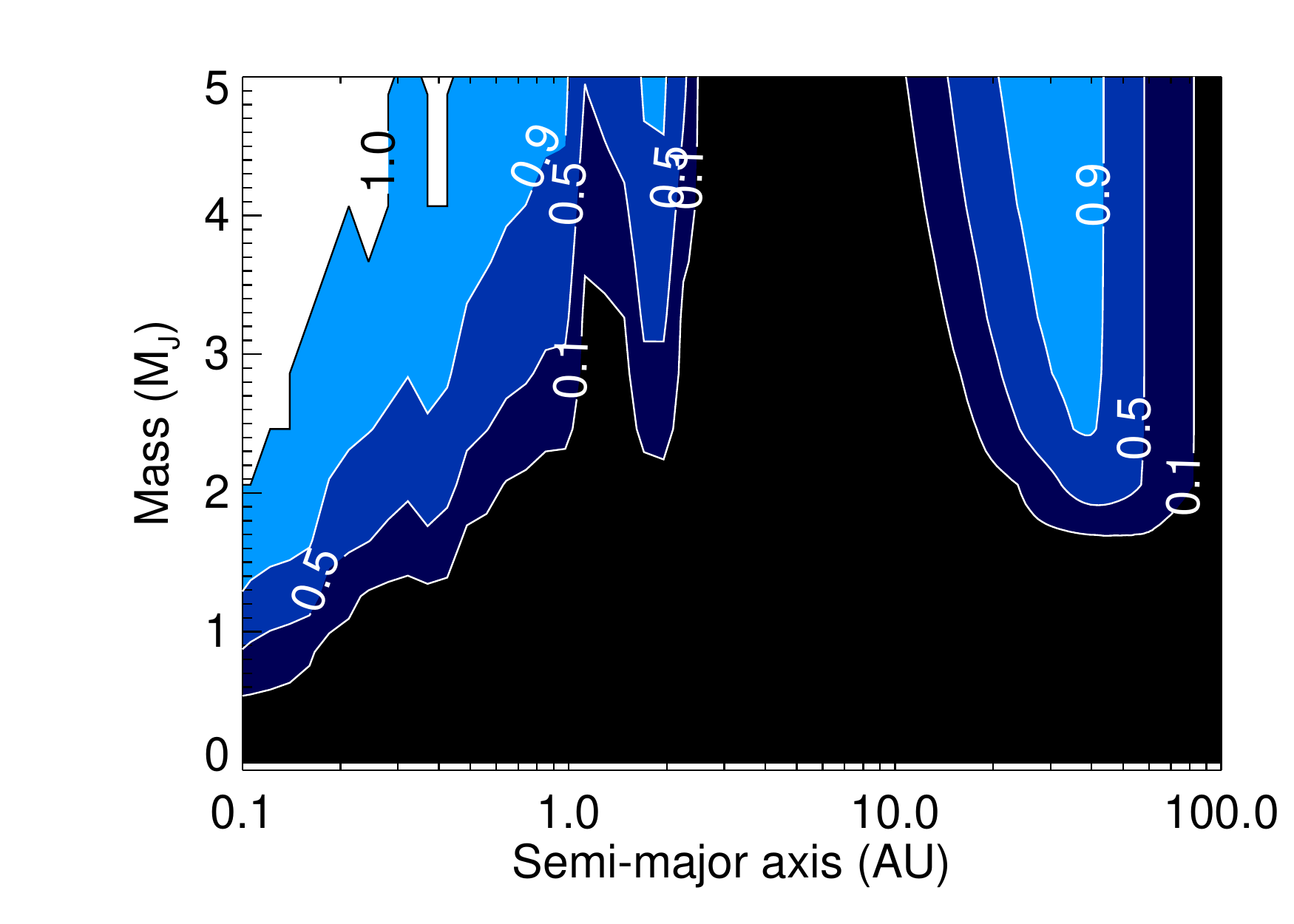}
\caption{Comparison of \texttt{MESS2} detection probabilities when removing a 2-planet Keplerian model from the RV (left; assuming the candidate planet HD113337~c is real) or only a 1-planet Keplerian model (right; assuming only the confirmed planet HD113337~b). This simulation was made in the case of the young age solution and for an inclination within 10 to 30\degr. Note that the {\it y}-scale is zoomed on the 0-5 \Mjup~range if compared to Fig.~\ref{fig:MESS2} (the left plot corresponds to the same simulation as the top left plot of Fig.~\ref{fig:MESS2}).}
\label{fig:MESS2_zoom}
\end{figure*}

We applied \texttt{MESS2} to the HD113337 \sophie~RV data set detailed in BO19+ (Sect.~\ref{sect:system}). We mainly used the RV corrected from the 2-planet Keplerian fit performed by BO19+, assuming that HD113337~c is an actual planetary companion. The RV analysis within \texttt{MESS2} relies on the periodogram-based Local Power Analysis method \citep[LPA,][]{meunier12}. Regarding the LBTI data, we used the contrast 2-D maps converted into masses considering the two possible age solutions of 20~Myr and 1~Gyr derived from the evolutionary track analysis (Sect.~\ref{sect:param}). In addition to the two age solutions, we also considered two synthetic planet inclination distributions: first, a uniform distribution between 0\degr~and 90\degr~(\ie, no assumption on $i$); and second, a narrow inclination range between 10\degr~and 30\degr~(assuming that the synthetic planets orbit within the same plane as the resolved outer disk). Thus, we mainly tested four \texttt{MESS2} simulations. We show the deduced detection probability curves in Fig.~\ref{fig:MESS2}.\\

The \sophie~RV data set allows us to rule out any additional companion to the known GP(s) with a mass above $\sim$5~\Mjup~up to 9~au if considering a system inclination around 25\degr, and still 90\%~of them if considering a uniform inclination distribution. We remind that we used RV residuals of a 2-planet Keplerian fit, explaining this good sensitivity. We overplotted the true masses (or minimal masses) of the known companions on Fig.~\ref{fig:MESS2} for the 25\degr~and uniform inclinations, respectively (top and bottom plots). The sudden sensitivity gap at $\sim$8~au is explained by the current impossibility to simulate the RV signal of synthetic planets with orbital periods longer than twice the RV data set time span within \texttt{MESS2} \citep{bonnefoy18}. We are not able to fully ``bridge'' the gap between the RV and imaging separation domains in any of the four simulations. Understandably, the detection probabilities provided by the LBTI images are best when considering the young age solution and close to pole-on inclination (top left plot). In this case, $\sim$100\%~of additional companions above $\sim$10~\Mjup~are excluded between $\sim$25 and $\sim$35~au, and $\sim$90\%~between $\sim$15 and $\sim$45~au. Increasing the age to 1~Gyr or assuming no hypothesis on the system inclination significantly reduce our sensitivity to companions within the separation domain covered by the LBTI. Compared to our ``classical'' (1-D) mass detection limits (Fig.~\ref{fig:lbti_limdets}), our \texttt{MESS2} detection probabilities show a decreased companion sensitivity within the separation range covered by the LBTI images. This can be expected as we computed the former 1-D detection limits as if assuming the HD113337 system was seen face-on (\ie~the most favourable case for imaging companions). We emphasize that we used only one epoch of observation regarding the imaging data within our \texttt{MESS2} analysis. A way to increase the sensitivity to companions within the imaging separation domain is to combine multiple high-contrast images acquired at different epochs \citep[see \eg][]{lannier17,lagrange18}.\\

We additionally tested the impact of assuming the presence of only one RV-detected GP in the HD113337 system (HD113337~b, confirmed planet) on our \texttt{MESS2} detection probabilities. To do so, we used the \sophie~RV data corrected from the 1-planet Keplerian model corresponding to planet b only, while keeping the longer-term RV variability (see more details in BO19+). We performed only one simulation, considering the young age solution and an inclination between 10\degr~and 30\degr. In this case, the detection probabilities at the shortest separations are significantly degraded (Fig.~\ref{fig:MESS2_zoom}), increasing the gap between the separation ranges covered by RV and direct imaging. This is expected as the computation of the RV detection probabilities within \texttt{MESS2} is based on the analysis of the RV Lomb-Scargle periodogram \citep{meunier12,lannier17}.

\section{Conclusion}\label{sect:conclu}

We combined different techniques to explore and bring constraints on various aspects of the HD113337 system. New optical long-based interferometric measurements allowed us to measure the linear radius of HD113337 with a precision better than 3\%. By using the new {\it Gaia} DR2 parallax and computing the star's bolometric flux, we were able to derive two very distinct isochronal age solutions for the system. The first (young) solution corresponds to an age of $\sim$14-21~Myr, while the second (old) one to an age of $\sim$0.8-1.7~Gyr. However, as often with such age degeneracies, we could not definitely settle the question of HD113337 age. For the first time, we were able to (partially) resolve HD113337 outer debris disk and to model its radius, its radial extension and, very interestingly, its inclination. We also found hints of possible inner disk components, which would be in agreement with previous SED studies. Next, we took new high-contrast images of the system's outer environment with the LBTI imager. We used both the deduced contrast limits and previous RV data to explore the complete GP-type companion (mass, separation) range up to 80-100 au from HD113337. At the same time, we took advantage of the age solutions and disk inclination value that we found to characterize the corresponding sensitivity to companions. Interestingly, this allowed us to deduce hints of the possible true masses of the HD113337~b confirmed GP and of the candidate HD113337~c. Furthermore, we were able to bring the first constraints on the presence of additional undetected companions at larger separations using the \texttt{MESS2} tool.\\

While it was not the main topic of this study, an important issue with the HD113337 system is to determine once and for all if the candidate companion HD113337~c is a real one (see BO19+). At this time, we are carrying out an additional long-term RV monitoring of HD113337 with the \sophie~spectrograph to increase our RV time span. This could allow us to remove the ambiguity of the RV and spectral line profile long-term signals. Another compelling possibility is to combine our RV data with astrometric data from \hipp~and/or {\it Gaia}. The tentative second GP that might orbit around HD113337 is an ideal target within this context. Furthermore, the RV~+~astrometric data combination would bring more constraints on the system's inclination and might even allow to confirm if the GP(s) orbit within the same plane as the outer debris disk. If present, and if seen inclined, HD113337~c would be a very massive planet, most probably formed through core-accretion while close at the same time of the commonly considered mass boundary between GP and brown dwarf companions. Finally, the combined RV~+~imaging analysis that we carried out in this study can be extended by using multi-epoch high-contrast imaging data, which would allow to fully ``close the gap'' with the RV sensitivity domain. HD113337 has already been included in several high-contrast imaging surveys. To conclude, we consider that HD113337 constitutes an exciting and rich system to further explore. It could make for a useful contribution for both stellar physics (with regard to stellar age determination), GP formation and evolution as a function of stellar properties, possibly multi-component debris disk studies, and planetary-disk interactions.

\begin{acknowledgements}
This work is based upon observations obtained with the Georgia State University Center for High Angular Resolution Astronomy Array at Mount Wilson Observatory. The CHARA Array is supported by the National Science Foundation under Grants No. AST-1211929 and AST-1411654. This work was supported by the Programme National de Physique Stellaire (PNPS) of CNRS/INSU co-funded by CEA and CNES. We acknowledge financial support from LabEx OSUG@2020 (Investissements d'avenir - ANR10LABX56). K.Y.L.S. acknowledges the partial support from the NASA grant NNX15AI86G. This work has made use of data from the European Space Agency (ESA) mission {\it Gaia} (\url{https://www.cosmos.esa.int/gaia}), processed by the {\it Gaia} Data Processing and Analysis Consortium (DPAC, \url{https://www.cosmos.esa.int/web/gaia/dpac/consortium}). Funding for the DPAC has been provided by national institutions, in particular the institutions participating in the {\it Gaia} Multilateral Agreement. This research has made use of the SIMBAD and VIZIER databases\footnote{Available at: \url{http://cdsweb.u-strasbg.fr/}} at the CDS, Strasbourg (France), and of electronic bibliography maintained by the NASA Astrophysics Data System (ADS) system. It has made use of the Jean-Marie Mariotti Center \texttt{Aspro} service\footnote{Available at: \url{http://www.jmmc.fr/aspro}}. We would like to thank our anonymous referee for the appreciative comments.
\end{acknowledgements}

\bibliographystyle{aa}
\bibliography{bibHD113337_2}

\end{document}